\PassOptionsToPackage{table,xcdraw}{xcolor}
\documentclass[sigconf]{acmart}
\usepackage{pifont}
\usepackage{colortbl}
\definecolor{lightgreen}{RGB}{204,255,204}
\definecolor{lightred}{RGB}{255,204,204}
\definecolor{lightgrey}{RGB}{224,224,224}
\usepackage{float}
\usepackage{enumitem}
\usepackage{multirow}
\usepackage{makecell}
\usepackage{graphicx}
\usepackage{subcaption}
\newcommand{\mxj}[1]{{\color{black} #1}}

\definecolor{TellingColor}{RGB}{94,197,186}
\definecolor{SellingColor}{RGB}{233,170,114}
\definecolor{ParticipatingColor}{RGB}{217,124,97}
\definecolor{DelegatingColor}{RGB}{149,163,201}
\definecolor{BaselineColor}{RGB}{163,163,163}
\definecolor{applegrey}{RGB}{99, 99, 102}

\copyrightyear{2026}
\acmYear{2026}
\setcopyright{cc}
\setcctype{by}
\acmConference[CHI '26]{Proceedings of the 2026 CHI Conference on Human Factors in Computing Systems}{April 13--17, 2026}{Barcelona, Spain}
\acmBooktitle{Proceedings of the 2026 CHI Conference on Human Factors in Computing Systems (CHI '26), April 13--17, 2026, Barcelona, Spain}
\acmPrice{}
\acmDOI{10.1145/3772318.3790551}
\acmISBN{979-8-4007-2278-3/2026/04}

\begin{document}

\title[Designing and Evaluating Strategies for LLM Agents to Advance Knowledge Co-Construction in Asynchronous Online Discussions]{"Shall We Dig Deeper?": Designing and Evaluating Strategies for LLM Agents to Advance Knowledge Co-Construction in Asynchronous Online Discussions}

\author{Yuanhao Zhang}
\email{yzhangiy@connect.ust.hk}
\affiliation{%
  \institution{Hong Kong University of Science and Technology}
  \city{Hong Kong}
  \country{China}
}

\author{Wenbo Li}
\email{wli55@ncsu.edu}
\affiliation{%
  \institution{North Carolina State University}
  \city{Raleigh}
  \country{United States}
}

\author{Xiaoyu Wang}
\email{xwangij@connect.ust.hk}
\affiliation{%
  \institution{The Hong Kong University of Science and Technology}
  \city{Hong Kong}
  \country{China}
}

\author{Kangyu Yuan}
\email{kyuanaf@connect.ust.hk}
\affiliation{%
  \institution{The Hong Kong University of Science and Technology}
  \city{Hong Kong}
  \country{China}
}

\author{Shuai Ma}
\email{mashuai@iscas.ac.cn}
\affiliation{%
  \institution{Institute of Software, Chinese Academy of Sciences}
  \city{Beijing}
  \country{China}
}

\author{Xiaojuan Ma}
\email{mxj@cse.ust.hk}
\affiliation{%
  \institution{Hong Kong University of Science and Technology}
  \city{Hong Kong}
  \country{China}
}

\renewcommand{\shortauthors}{Zhang et al.}

\begin{abstract}
Asynchronous online discussions enable diverse participants to co-construct knowledge beyond individual contributions. This process ideally evolves through sequential phases, from superficial information exchange to deeper synthesis. However, many discussions stagnate in the early stages. Existing AI interventions typically target isolated phases, lacking mechanisms to progressively advance knowledge co-construction, and the impacts of different intervention styles in this context remain unclear and warrant investigation. To address these gaps, we conducted a design workshop to explore AI intervention strategies (task-oriented and/or relationship-oriented) throughout the knowledge co-construction process, and implemented them in an LLM-powered agent capable of facilitating progression while consolidating foundations at each phase. A within-subject study (N=60) involving five consecutive asynchronous discussions showed that the agent consistently promoted deeper knowledge progression, with different styles exerting distinct effects on both content and experience. These findings provide actionable guidance for designing adaptive AI agents that sustain more constructive online discussions.

\end{abstract}





\maketitle

\section{Introduction}

\begin{figure*}[h]
  \centering
  \includegraphics[width=0.6\linewidth]{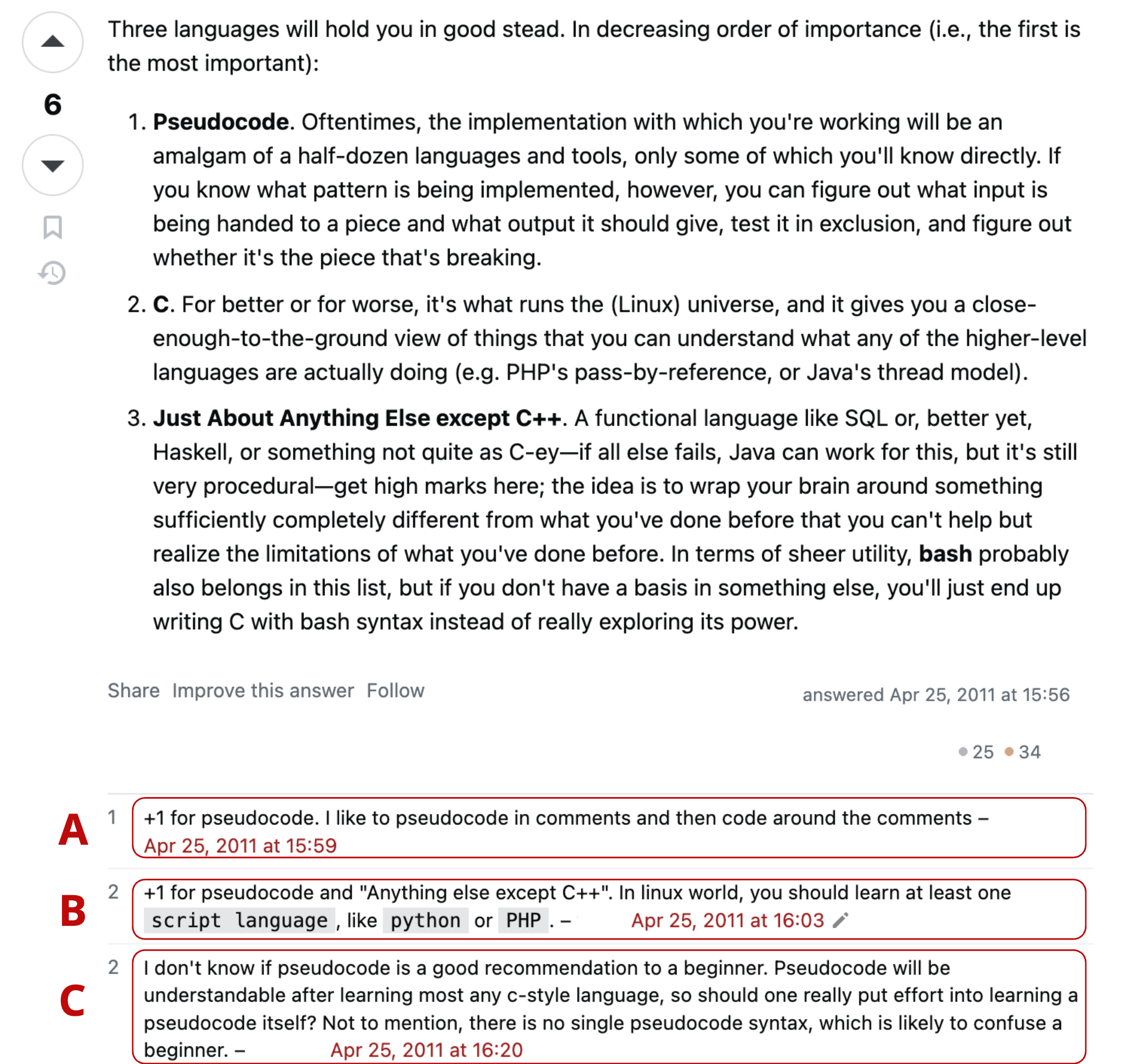}
  \caption{The screenshot shows a Stack Exchange discussion thread responding to the question “What language for starting on Linux?”, where the accepted answer (top) recommends learning pseudocode, C, and any scripting language except C++. Comments A, B, and C co-construct knowledge by complementing, elaborating, or challenging this suggestion. Comment A agrees with and expands on the value of pseudocode. Comment B reinforces the recommendation while adding specific language examples such as Python and PHP. In contrast, Comment C challenges the practicality of learning pseudocode for beginners and raises concerns about its lack of standard syntax. Together, these interactions illustrate how community members collaboratively refine and contest knowledge claims, contributing to a richer collective understanding.}\label{fig:intro-example}
\end{figure*}

Asynchronous online discussion platforms (e.g., Reddit \cite{reddit_2005}, Discourse \cite{discourse}) have become essential venues for knowledge co-\hspace{0pt}construction \cite{dubovi_empirical_2020, weinberger_framework_2006, yogyakarta_state_university_analysis_2023}, a process that enables participants from diverse backgrounds to collaboratively examine, extend, and refine one another’s ideas, thereby enhancing shared understanding of complex issues across various disciplines \cite{zhang_understanding_2023, jones_rscience_2019, lucas_assessing_2014} (Figure~\ref{fig:intro-example}). 
Compared with individually held knowledge, such collective processes integrate multiple perspectives, reconcile divergent viewpoints, and generate more robust and comprehensive understandings of a subject \cite{gunawardena_analyzing_nodate, dubovi_empirical_2020, he_engage_2024}.  
Ideally, knowledge co-construction unfolds through a staged interactive progression, evolving from initial idea sharing (\textit{initiation}), building on and enriching each other’s contributions (\textit{exploration}), explicitly addressing divergent perspectives to reach shared understanding (\textit{negotiation}), to finally synthesizing and reflecting jointly created knowledge (\textit{co-construction}) \cite{onrubia_strategies_2009, gunawardena_analysis_1997, stahl2013model}.
\mxj{
While surface-level exchanges suffice for socializing or simple information sharing, it is critical for discussions centered on complex or open-ended topics to achieve the necessary depth \cite{de2006content}.
}
However, prior empirical research indicated that many such asynchronous discussions remain concentrated in the early, surface-level stages, with limited progression toward more integrative phases \cite{dubovi_empirical_2020, lucas_assessing_2014, nguyen_facilitating_2023}. 
For users, shallow exchanges deprive contributors of meaningful uptake -- leaving their inputs unexamined, unelaborated, and unintegrated -- while consumers struggle to obtain coherent explanations or reconciled viewpoints, instead encountering fragmented comments \cite{lee2005asynchronous, black2005use, de2019expect}.
For platforms, such stagnation undermines their value as sustainable knowledge repositories, producing piecemeal discussions rather than synthesized collective knowledge \cite{de2019expect}.
Therefore, the need arises for dynamic support for users of these platforms to achieve sustained and meaningful collective knowledge building as their discussions unfold \cite{li2023research}.

Advances in AI technology have spurred the deployment of agents beyond one‑to‑one interactions to multi‑party discussions, where they employ various strategies to support and shape group dynamics \cite{tan_systematic_2022, do2022should}. 
One type of agents aims at interactional coordination and behavior management, such as ensuring balanced participation \cite{bagmar2022analyzing, kim2020bot, do2023inform}, emotion regulation \cite{benke2020chatbot, peng2019gremobot, wang2023reprompt}, and discussion pacing \cite{kim2021moderator, shin2023introbot}. 
They helped smooth the discussion process, but paid limited attention to the depth and breadth of content development. 
Another line of automated interventions focuses directly on supporting knowledge co-construction, for instance by persistently challenging claims \cite{chiang2024enhancing, ma2025towards} or prompting clarifications \cite{cai_advancing_2024, ito_agent_2022, hadfi_argumentative_2021}. 
These interventions typically focus on content issues within a particular phase, without a broader, process-orchestrated emphasis on fostering the progression across phases.  
Thus, the design of strategies for meaningfully advancing knowledge co-construction remains largely underexplored.  

Previous studies suggest that humans exercise different styles of interventions to promote knowledge co-construction, \mxj{generally distinguished between task-oriented approaches (providing guidance to advance knowledge-building \cite{wang2025social, nguyen2023role}) and relationship-oriented approaches (contributing alongside others to foster closer and more equal collaboration \cite{chiang2024enhancing, zhang2021ideal}).}
These facilitation styles are likely to shape group dynamics, motivation, and knowledge-sharing outcomes in distinct ways \cite{raza2018impact, del2025hersey}. 
While these findings concern human facilitation, prior studies indicate that agents engaging in multi‑party interactions with different intervention styles can elicit varied reactions from human participants \cite{kim2024engaged, xu2013designing}.
These insights motivate our investigation into how different AI intervention styles may affect the knowledge co-construction dynamics around content development and participant experiences during asynchronous online discussions.

To this end, we first conducted a design workshop with 12 participants (nine active contributors to online knowledge discussions and three AI designers) to co-design the intervention strategies an AI member can take to advance knowledge development across different co-construction phases. The strategies were designed in telling, selling, participating, and delegating styles, respectively. Participants first identified the sufficiency criteria for each phase, then brainstormed and calibrated strategies that an AI agent with a given style could employ to achieve these criteria. After refinement and synthesis of these insights, we further developed an LLM-powered agent to progressively advance online knowledge co-construction, ensuring robust foundations at each phase. The agent can analyze each comment's characteristics within the evolving discussion, monitor phase-level dynamics, and operationalize workshop-derived, phase-specific interventions in different styles. 
With the agent, we proceeded to addressing the following research questions:

\begin{description}
    \item[\hspace{0.6em}RQ1] How do the AI intervention styles influence the progression of knowledge co-construction in asynchronous online discussions?
    \item[\hspace{0.6em}RQ2] How do the intervention styles shape human perceptions of the AI agent and their user experience?
    \item[\hspace{0.6em}RQ3] How do the intervention styles affect human-human interaction in asynchronous online discussions?
\end{description}


To answer these questions, we conducted a within-subject study (N = 60) where each participant engaged in five consecutive discussion threads, each involving six participants, under five conditions (four AI intervention styles and one human-only baseline).
Mixed-methods analyses -- including thread-level metrics, in-task surveys, and interviews -- revealed that compared with the baseline, agents employing telling, selling, and participating styles had the potential to advance the knowledge co-construction into deeper phases and to establish a stronger foundation in the \textit{exploration} phase, with each style exhibiting distinct strengths and limitations in driving such advancements. These three styles were also perceived more favorably than the delegating style, with the participating style receiving the highest appreciation. Participants also reported that different styles influenced human–human interactions in distinct ways.

The contributions of our work are threefold. First, we design an actionable set of phase‑tailored intervention strategies in different styles for AI agents to advance knowledge co-construction as active discussion participants. 
Second, our mixed-methods evaluation provides empirical insights into the impacts of various AI intervention styles on knowledge co-construction and participants' experiences.
Third, we distill key design implications to guide the development of adaptive AI agents for multi‑party interactions.

\section{Related Work}

\subsection{Knowledge Co-Construction in Asynchronous Discussion Platforms}

Asynchronous discussion forums (e.g., Reddit \cite{reddit_2005}, Discourse \cite{discourse}, Stack Exchange \cite{stackexchange_2019}) have become crucial online platforms where millions of users with diverse experiences and expertise engage in knowledge seeking and information sharing \cite{ august_explain_2020, du2022enhancing}. Unlike ephemeral real-time conversations, asynchronous forums transform user interactions into persistent knowledge repositories \cite{de2024assessing, erickson2005persistent}. Posts and replies, preserved as textual records, allow continuous access, quotation, and iterative elaboration, thus establishing a cumulative community knowledge base \cite{erickson2005persistent}. Additionally, asynchronous forums offer temporal flexibility, enabling participants to thoughtfully contribute at their own pace \cite{alrushiedat2012anchored}. These characteristics make asynchronous forums particularly effective for harnessing the ``wisdom of crowds'' and supporting knowledge co-construction.

Knowledge co-construction refers to a collaborative process wherein individuals not only share information but also collectively refine, negotiate, and expand each other's ideas, fostering collective understanding beyond individual contributions \cite{dubovi_empirical_2020, kimmerle_learning_2015, yogyakarta_state_university_analysis_2023}. It can be conceptualized as a progressive, multi-phase process, evolving from superficial information sharing toward deeper integration \cite{gunawardena_analysis_1997, onrubia_strategies_2009, stahl2013model}. Among existing frameworks describing this process, the Interaction Analysis Model (IAM) \cite{gunawardena_analysis_1997} is widely recognized. IAM characterizes knowledge co-construction as five sequential phases: (1) sharing ideas, (2) exploring dissonance, (3) negotiating meaning, (4) testing tentative knowledge syntheses, and (5) applying newly constructed knowledge.
Lucas et al. \cite{lucas_assessing_2014} empirically assessed IAM and suggested that higher phases may require reconsideration or consolidation. In line with this perspective, Onrubia et al. \cite{onrubia_strategies_2009} proposed a simplified four-phase analytical model inspired by IAM, comprising phases of (1) initiation, (2) exploration, (3) negotiation, and (4) co-construction. Stahl’s collaborative knowledge-building model \cite{stahl2013model} also shares conceptual similarities, emphasizing the progression from publicly stated ideas through argumentation development, negotiation toward shared understanding, and eventually the objectification of collaboratively generated knowledge. Building upon these empirically and theoretically validated models, this paper employs a refined four-phase coding scheme (detailed in Section \ref{sec: 3.1}) to analyze knowledge co-construction in our study. 

Empirical studies analyzing asynchronous knowledge co-\hspace{0pt}construction using these frameworks consistently observed that interactions predominantly occur at superficial levels, with significantly fewer interactions reaching the deeper phases \cite{lu_knowledge_2006, zhu_beyond_2023, lucas_assessing_2014}. For instance, Dubovi et al. \cite{dubovi_empirical_2020}, employing IAM to analyze comments on YouTube videos, reported that most interactions remained at initial sharing and exploration phases. McLoughlin et al. \cite{mcloughlin2000cognitive} found similar patterns in their analysis of university students' interactions in online learning forums.
Such stagnation limits contributors’ opportunities to have their ideas examined and built upon, discouraging sustained engagement \cite{black2005use, lee2005asynchronous}; leaves consumers with fragmented and sometimes contradictory information that hinders coherent understanding \cite{dubovi_empirical_2020, de2024assessing}; and reduces platforms’ value as repositories of synthesized, durable knowledge \cite{de2019expect}.
To address this gap, our work explores the potential of AI agents to engage in multi-party interactions as active discussion participants, strategically fostering more advanced phases of knowledge co‑construction.




\subsection{\mxj{Interventions to Support Knowledge Co-construction}}
\label{sec: 2.2}

\mxj{
Extensive research in the learning sciences demonstrates that human intervention is critical for fostering productive knowledge co-construction \cite{kaendler2015teacher, van2019systematic}, which generally fall into two categories. The first is instructional intervention, where facilitators actively guide the discussion with a strong task focus. For instance, teachers can prompt students to elaborate on their explanations \cite{webb2008role} or ask specific follow-up questions to deepen the discussion \cite{webb2009teacher},  thereby driving the group toward learning goals. The second adopts a more participatory approach, characterized by egalitarian, teammate-like relationships. In this mode, teachers can act as "role models" by demonstrating example behaviors \cite{tanis2014beter, van2019systematic}, or, as proposed by King \cite{king1997ask}, peers can assess and tutor one another to foster mutual knowledge construction.
}

\mxj{With advances in Natural Language Processing (NLP), AI agents have increasingly been deployed to support multi-party interactions.
A significant strand of this research focuses on agents acting as coordinators, emphasizing interaction management over substantive knowledge building.
For instance, agents like IntroBot \cite{shin2023introbot} and BlahblahBot \cite{shin2021blahblahbot} focus on social lubrication (e.g., icebreakers), while others like GroupfeedBot \cite{kim2020bot} regulate participation rates or manage group emotion \cite{benke2020chatbot, peng2019gremobot}. While effective for maintaining productive dynamics, these "coordinator" agents mainly manage the flow of conversation without enhancing the quality of the knowledge being constructed.}


\mxj{
To directly support knowledge co-construction, more recent work has begun to assume the intervention roles traditionally held by humans. Mirroring the strategies above, these AI interventions generally adopt two distinct approaches. The first positions the AI as a facilitator that directs the knowledge-building process. For instance, some agents \cite{cai_advancing_2024, ito_agent_2022, nguyen2023role} post follow-up questions to prompt users to refine and reflect on their prior statements. Other agents steer groups toward deeper collaboration by delivering timely prompts (e.g., JIA \cite{doherty2025piecing}) or enriching discussions with context-relevant informational suggestions (e.g., PAPERPING \cite{wang2025social}). The second category positions the AI as a teammate \cite{zhang2021ideal}, which actively contribute task knowledge rather than providing guidance. For instance, Chiang et al. \cite{chiang2024enhancing} developed a "devil’s advocate" agent that provides arguments from a position opposing the majority opinion.
However, these interventions typically address isolated conversational moments, without a process-orchestrated paradigm to adapt their support to the evolving stages of knowledge co-construction. This study addresses this gap by conducting a \hyperref[design-workshop]{Design Workshop} to develop phase-specific AI intervention strategies and a narrative literature review to derive empirically grounded design requirements for systematically advancing the depth and quality of collective knowledge in asynchronous discussions. 
}





\subsection{Frameworks of Intervention in Group Settings}

Designing AI intervention strategies for multi-party interaction can be informed by models validated in human group contexts \cite{govers2024ai, chen2025maintaining}. Task-Relationship (TR) framework depicts two distinct behavioral dimensions commonly used to categorize human intervention styles \cite{stogdill1974handbook}: \textbf{task} orientation, which emphasizes providing direction and achieving goals, and \textbf{relationship} orientation, which emphasizes fostering interpersonal support and active collaboration through shared responsibility. Together, these dimensions form a descriptive framework characterizing most observed intervention behaviors in group settings \cite{forsyth2014group}. 
Several influential models further conceptualized specific styles within the space defined by these two dimensions  \mxj{through a quadrant logic (e.g., Managerial Grid Model\cite{blake1964managerial}, Situational Leadership Model (SLM) \cite{hersey1979situational, fiedler1967theory}).}

\mxj{
TR framework also effectively characterize the landscape of AI interventions for knowledge co-construction reviewed in Section \ref{sec: 2.2}. Specifically, facilitator-like agents \cite{doherty2025piecing, wang2025social, nguyen2023role}, which focus on providing guidance and structuring the discussion to foster knowledge building, are more \textbf{task}-orientated. The teammate-like roles \cite{chiang2024enhancing, zhang2021ideal}, which build rapport and equal relationship by acting as "lead-by-example" peers (i.e., actively contributing knowledge alongside others for demonstration), are more \textbf{relationship}-oriented.
Due to this conceptual alignment, this work contextually adapts the TR framework to define AI intervention styles. Drawing on the practices of prior models \cite{hersey1979situational, fiedler1967theory}, we also apply the quadrant logic-mapping the intersection of these two dimensions—to systematically explore how different combinations of guidance (task) and peer-like participation (relationship) influence knowledge co-construction.
}

\mxj{Research indicates that intervention style can shape group outcomes. 
For example, Lu et al. \cite{lu_knowledge_2006} found that adopting a peer-like role—where teachers actively contribute to the discussion alongside students—led to higher-quality knowledge construction than merely facilitating. 
In contrast, Raza et al. \cite{raza2018impact} reported greater group achievement from more task-oriented interventions.}
While these findings concern human intervention, prior studies also suggest that participants are sensitive to the styles used by AI agents in group settings and may respond differently depending on the style \cite{kim2024engaged, xu2013designing}. 
Even though AI agents act as equal members in asynchronous discussions, they become a temporary, ad hoc leader when trying to engage other members to advance knowledge co-construction. Like humans, AI agents have the choice to prioritize task completion (i.e., driving the group toward synthesizing collective knowledge) and/or relationship (i.e., nurturing mutual engagement and shared ownership of the knowledge), yet their impacts in knowledge building are underexplored. This gap motivates our investigation into how AI agents, when adopting these intervention styles, influence both the dynamics of knowledge co-construction and participants’ experiences in asynchronous forums.

\section{Design and Implementation}
\label{sec: 3}
This section illustrates the design and implementation of AI intervention strategies. Based on theoretical frameworks, we conducted a design workshop to identify agent strategies for advancing knowledge co‑construction across different styles. Insights from the workshop informed the development of an LLM‑based agent. 

\subsection{Theoretical Framework}
\label{sec: 3.1}
The design of strategies across different phases and styles requires a theoretical grounding in process models of knowledge co-\hspace{0pt}construction and established frameworks of intervention styles. 
Drawing on IAM \cite{gunawardena_analysis_1997}, Onrubia et al. \cite{onrubia_strategies_2009}, and Stahl’s collaborative knowledge‑\hspace{0pt}building model \cite{stahl2013model}, we adopt a refined four‑phase coding scheme to examine knowledge co‑construction in asynchronous discussions: 
\begin{enumerate}[label=\textbf{Phase\arabic*}, leftmargin=4em]
    \item \textbf{Initiation}: Introducing novel perspectives without engaging in interactive discourse with others.
    
    \item \textbf{Exploration}: Building upon or complementing prior contributions with their own information.
    
    \item \textbf{Negotiation}: Explicitly addressing the inconsistencies identified in earlier discussions.

    \item \textbf{Co-construction}: Collaboratively integrating negotiation insights into a coherent synthesis, and reflecting on the jointly created knowledge.
\end{enumerate}

\mxj{
Note that while these phases model an ideal progression of knowledge construction, real-world discourse is often iterative rather than strictly linear. Discussions frequently cycle between phases—particularly exploration and negotiation—as new participants enter.
}

\mxj{For intervention styles, we apply quadrant logic to the TR framework \cite{stogdill1974handbook}, mapping the intersection of \textbf{task} and \textbf{relationship} dimensions. We adapt the resulting four styles \footnote{\mxj{Our definitions are a contextual adaptation of the TR framework; similarity to SLM nomenclature does not imply an adopting its original definitions.}} to the context of knowledge co-construction}:

\begin{itemize}
\item \textbf{Telling}: High-task, low-relationship -- the agent provides explicit instructions to guide knowledge co-construction.
\item \textbf{Selling}: High-task, high-relationship -- the agent suggests directions for knowledge co-construction and actively persuades others to adopt them, fostering shared commitment to the emerging knowledge.
\item \textbf{Participating}: Low-task, high-relationship -- the agent reduces directive behavior, instead collaborating with the group and sharing responsibility for constructing knowledge.
\item \textbf{Delegating}: Low-task, low-relationship style -- the agent adopts a hands-off approach, granting autonomy to members while monitoring progress with minimal intervention.
\end{itemize}

\subsection{Design Workshop}
\label{design-workshop}
\mxj{
We conducted a design workshop to identify agent strategies for advancing knowledge co-construction. While the theoretical framework are grounded in prior literature, operationalizing these concepts requires capturing the situated practices of online communities. This participatory approach complements the literature by surfacing contributors' tacit knowledge, ensuring our design is ecologically valid and aligned with real-world dynamics.
}

\subsubsection{Participants and Procedure}
With IRB approval, we conducted a three‑hour online design workshop with 12 participants (four female, eight male), recruited via online advertisements, social media, and word-of-mouth. The group comprised three AI designers\mxj{—two of whom specialize in AI in Education with backgrounds in learning sciences—}and nine frequent contributors to online knowledge discussions (three contributed daily, four multiple times a week, and two at least once a week; \mxj{summarized in Table \ref{tab:participants_profile}}). 

\begin{table*}[htbp]
  \centering

  \renewcommand{\arraystretch}{1.2} 
  \begin{tabular}{ccccc}
    \toprule
    \textbf{ID} & \textbf{Gender} & \textbf{Primary Platform} & \textbf{Freq. of Contribution} & \textbf{Confidence Level} \\
    \midrule
    PC1 & Male   & Zhihu          & At least once a week  & 5 \\
    PC2 & Male & Reddit         & Multiple times a week & 5 \\
    PC3 & Female   & Zhihu          & Daily & 6 \\
    PC4 & Female & Quora          & Multiple times a week & 7 \\
    PC5 & Female   & Stack Exchange & Multiple times a week & 6 \\
    PC6 & Female   & Stack Exchange & Multiple times a week                 & 6 \\
    PC7 & Male & Zhihu          & At least once a week  & 4 \\
    PC8 & Male & Quora          & Daily                 & 6 \\
    PC9 & Male   & Reddit         & Daily                 & 5 \\
    \bottomrule

\end{tabular}
\caption{\mxj{Demographics and online behaviors of the active contributors in the workshop. Confidence level is self-reported on a 7-point Likert scale (1=Low, 7=High)}}.
\label{tab:participants_profile}
\end{table*}


The workshop was held via a video‑conferencing platform \cite{voovmeeting}, with a shared online whiteboard tool \cite{miro} for collaborative editing. It began with a 15‑minute briefing on the definitions of the four knowledge co‑construction phases and the four intervention styles, supplemented by concrete examples to ensure a shared theoretical foundation. Participants then engaged in three sequential activities:
\begin{itemize}
    \item \textbf{Activity 1: Defining Sufficiency Criteria for Each Phase (15 minutes)} Building on theoretical definitions, participants collaboratively specified the sufficiency criteria for each phase. They shared perspectives, discussed differences, and consolidated these into an agreed set of outcomes, which served as design targets for subsequent activities. The results were documented in the shared spreadsheet.
    \item \textbf{Activity 2: Brainstorming Phase‑Specific Intervention Strategies (60 minutes)} Participants were divided into three groups (each comprising one AI designer and three frequent contributors), each assigned one intervention style—\hspace{0pt}Telling, Selling, or Participating. In breakout rooms with a facilitator, groups brainstormed agent strategies that could support phase transitions and achieve the sufficiency criteria from Activity 1. Contributions could be abstract strategy descriptions or concrete message examples, informed by participants’ own discussion experiences. Outputs were recorded in the shared document.  We excluded the Delegating style since its low‑task, low‑relationship orientation inherently requires minimal agent intervention, allowing it to be implemented through simple, predefined phase‑opening prompts without the need for extensive co‑design.
    \item \textbf{Activity 3: Consolidating and Refining Representative Strategies (75 minutes)} All participants reconvened to exchange and refine ideas generated in Activity 2, guided by a facilitator. New suggestions were welcomed during this stage. The process resulted in a final set of representative intervention strategies for each style (except delegating) across all four phases.
\end{itemize}

The workshop was audio‑recorded and transcribed, and all participant\hspace{0pt}-generated documents were retained. Two researchers independently analyzed the transcripts and artifacts, iteratively discussing and refining both the sufficiency criteria (from Activity 1) and the strategies (from Activities 2–3) to ensure that they accurately represented participants’ intended meanings and design rationales. Discrepancies were resolved through discussion until consensus was reached.

\begin{figure*}[h]
  \centering
  \includegraphics[width=\linewidth]{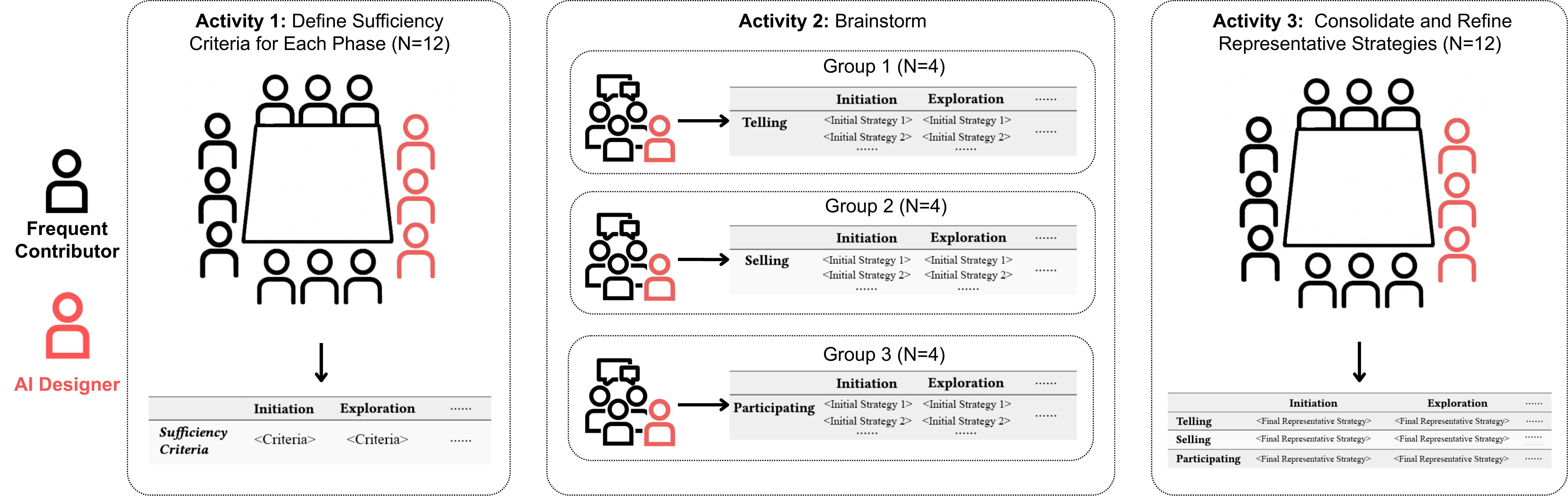}
  \caption{\mxj{Activities of the workshop. Sufficiency criteria for each phase were defined by the full cohort (N = 12), while representative intervention strategies for each agent style were first synthesized in subgroups (N = 4) and then validated by all participants (N = 12)}}\label{fig:workshop-pipeline}
\end{figure*}

\subsubsection{Summary of Findings}


\mxj{Table \ref{tab:workshop} summarizes the sufficiency criteria for each phase and the representative intervention strategies for each agent style. Given the collaborative structure of the workshop, results reflect group consensus rather than individual frequencies: sufficiency criteria were established by the full cohort ($N=12$), while strategies were synthesized in subgroups ($N=4$) and subsequently validated by all participants ($N=12$). Below, we detail how these outcomes manifest across the four phases.}

\begin{table*}[htbp]
\centering
\includegraphics[width=\textwidth]{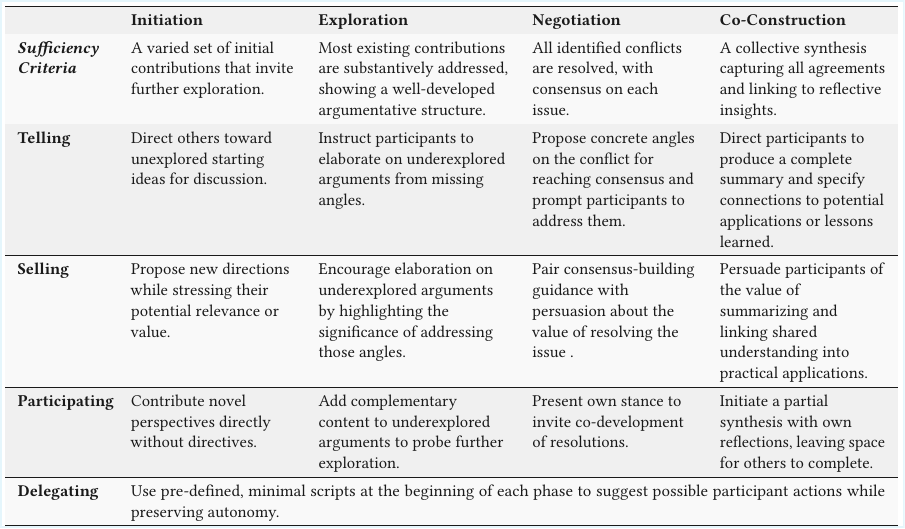}
\caption{Summary of findings from the design workshop, including the desired outcomes for each knowledge co-construction phase and the corresponding intervention strategies in different styles. The delegating-style strategy was not derived from the workshop due to its minimal intervention nature.}
\label{tab:workshop}
\end{table*}

\textbf{Initiation:} 
The desired outcome of this phase is to cultivate a broad range of starting points, offering a foundation for subsequent interactions.
To support this, telling-style agents were envisioned as issuing directives that introduce novel starting ideas not yet present in the conversation, prompting others to discuss them. Selling-style agents would propose such ideas while emphasizing their potential relevance or value. Participating-style agents would contribute novel perspectives of their own.

\textbf{Exploration:} 
This phase focuses on expanding existing contributions, ensuring that most are substantively addressed. Workshop participants suggested assessing sufficiency by examining the completeness of the argumentative structure. Telling-style agents could identify underexplored arguments and instruct participants to elaborate on them from specific missing angles. Selling-style agents would encourage such elaboration by underscoring its potential significance. Participating-style agents extend these contributions themselves, contributing complementary content to foster further exploration.

\textbf{Negotiation:} 
In this phase, sufficiency is indicated when inconsistencies emerged earlier have been resolved, with consensus reached on each issue. 
\mxj{However, achieving full agreement may not always be the ideal outcome, such as when it risks suppressing healthy disagreement or forcing premature convergence. Therefore, a valid "consensus" in this context also encompasses states where distinct viewpoints are clearly clarified and acknowledged.}
Participants noted that agents could identify unresolved disagreements and stimulate discussion towards such resolution. Telling-style agents could propose concrete approach for reaching consensus and ask participants to engage accordingly. Selling-style agents would pair this guidance with persuasion about the value of resolving the issue (e.g., benefits, risks of deferral). Participating-style agents present their own stance as a starting point for engaging others in co-developing negotiated agreements.

\textbf{Co-Construction:} 
The final phase focuses on collaboratively synthesizing agreements reached and linking them to practical applications or reflective insights. 
\mxj{Participants identified the active process of synthesis—rather than just the resulting artifact—as the critical mechanism for solidifying consensus.
Consequently, success was defined as a collective synthesis where contributors explicitly articulate the agreed knowledge, supported by substantial reflection.}
In this phase, telling-style agents explicitly instruct participants to produce a holistic summary, specifying which issues to address and from which perspectives to reflect. Selling-style agents, in addition to providing guidance, advocate for the value of comprehensive summarizing and reflection. 
\mxj{Participating-style agents, acknowledging the potential labor burden, initiate a partial summary with their own reflections to scaffold the process, yet intentionally leave gaps for users to complete and enrich.}

\subsection{Agent Design and Implementation}

To evaluate the derived strategies, we developed an LLM agent that operationalizes them in asynchronous discussion contexts. Drawing on design requirements derived from our research motivation and relevant literature, we proposed an intervention logic to guide the agent’s behavior toward these objectives. We then designed the agent’s architecture, which integrates this logic into a coherent computational framework.

\subsubsection{Design Requirements}
\mxj{To derive design requirements (DRs) that are theoretically grounded and empirically relevant, we conducted a narrative literature review \cite{baumeister1997writing}. We employed a hybrid search strategy, combining keyword-based retrieval on Google Scholar (e.g., \textit{“knowledge co-construction,”} \textit{“collaborative knowledge building”}) with snowballing from seminal frameworks like IAM \cite{gunawardena_analysis_1997} and Stahl’s model \cite{stahl2013model}. We targeted empirical studies—ranging from observational analyses of discussion datasets to experimental evaluations of collaborative processes—that investigate the dynamics of knowledge construction. From this pool, we prioritized research situated in asynchronous contexts that identified barriers to effective co-construction. A total of 26 key papers were synthesized to formulate the following DRs:}
\begin{enumerate}[label=\textbf{DR\arabic*}, leftmargin=3em]
    \item \textbf{Progression Facilitation:} 
    Models of knowledge co-\hspace{0pt}construction (e.g., IAM \cite{gunawardena_analysis_1997}, Stahl’s model \cite{stahl2013model}) conceptualize it as a sequence of progressively demanding phases, yet empirical work of asynchronous discussions shows most contributions remain clustered in early stages.
    Participants typically lack explicit awareness of the discussion’s current phase, and sufficient skills to determine when and how to advance the discourse \cite{fischer2013toward, kollar2006collaboration}.
    Therefore, the agent should monitor the evolving discussion and deliver well-timed stimulus, \mxj{aiming to promote progression towards higher phases rather than forcing all discussions to reach the final co-construction phase.}
    
    \item \textbf{Phase Sufficiency:} 
    The agent should avoid advancing discussions prematurely, as each phase provides a necessary foundation for the next \cite{stahl2013model}. Progressing to later stages without fully developing the preceding ones risks undermining collaboration quality: without diverse initiation, later stages lack the breadth of perspectives needed for productive exchange, leading to narrow deliberation \cite{stahl2013model}; without thorough exploration, negotiation may devolve into shallow or emotional disputes rather than rational debate \cite{onrubia_strategies_2009}; without rigorous negotiation, synthesis is prone to superficial consensus, producing fragile rather than durable knowledge \cite{weinberger_framework_2006}.
    The agent should therefore ensure that each phase meets the sufficiency criteria -- defined in the design workshop -- before moving forward.

    \item \textbf{Agency Preservation:}  
    While well‑timed support can foster sustained participation \cite{pea2018social, stefanou2004supporting}, overly frequent or prescriptive interventions may erode autonomy, diminishing intrinsic motivation, ownership of ideas, and creative contributions \cite{ryan2000self, dillenbourg2002over}. The agent should therefore act selectively, offering only the necessary input to sustain productive dialogue while leaving participants sufficient space to steer the discourse and co‑construct knowledge on their own terms.
\end{enumerate}

Guided by \textbf{DR1} and \textbf{DR2}, we propose a high-level intervention logic for advancing knowledge co-construction in asynchronous discussions (Figure \ref{fig:architecture}). At each phase, the agent evaluates whether the discussion meets the sufficiency criteria defined for that stage. If underdeveloped, the agent applies targeted interventions until the desired outcomes are achieved. Once met, the agent initiates strategies for the next phase to foster phase transition, and repeats the process.
\mxj{Specifically, this process accommodates latecomers contributing earlier-stage ideas without suppressing them to enforce a strict forward progression.}





\subsubsection{Agent Architecture}

\begin{figure*}[h]
  \centering
  \includegraphics[width=\linewidth]{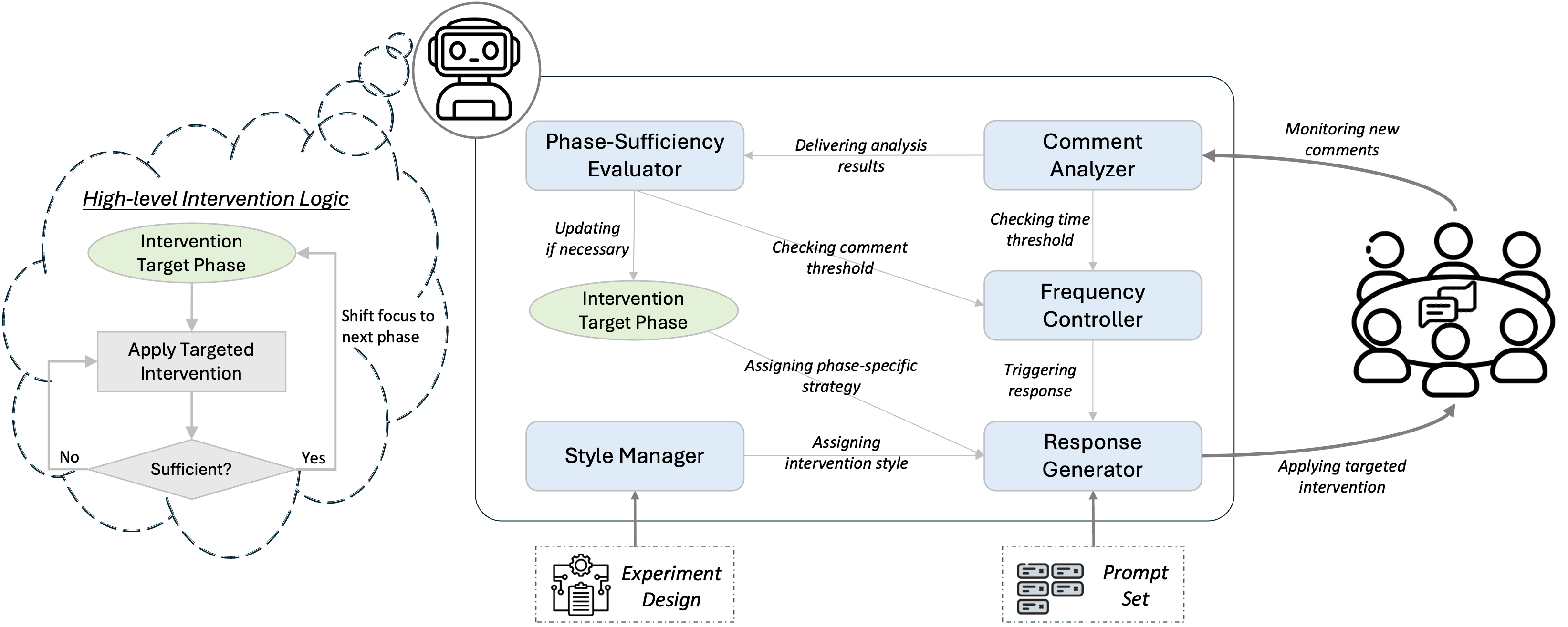}
  \caption{Architecture of the LLM agent implemented in our experiment, detailing its key components and operational logic. The architecture integrates high-level intervention logic to support targeted interventions in multi-party discussions, thereby facilitating the progressive advancement of knowledge co-construction.}\label{fig:architecture}
\end{figure*}

We implemented an LLM \footnote{\mxj{
During our experiment (August 2025), we used Gemini 2.5 Flash (hereafter “LLM”),  selected for its strong balance of reasoning, latency, and cost-efficiency. As our architecture relies on generic capabilities rather than model-exclusive features, we anticipate comparable performance across other SOTA models.}
} agent whose architecture comprises five core components: \textit{Comment Analyzer}, \textit{Frequency Controller}, \textit{Phase-Sufficiency Evaluator}, \textit{Style Manager}, and \textit{Response Generator} (Figure \ref{fig:architecture}). All numerical thresholds \mxj{(e.g., intervention intervals)} used in this section were empirically derived from pilot trials to ensure smooth discussion flow in our experimental setting. Technical details (e.g., algorithms, prompts, evaluations) are included in the supplementary material.

\textbf{Comment Analyzer} : 
This component continuously monitors incoming comments and processes them through its submodules to construct a structured understanding of the discussion, which serves as the basis for subsequent interventions.  

1. \textit{Phase Classifier} — Consistent with previous empirical approaches \cite{dubovi_empirical_2020, nguyen_facilitating_2023, zhu_beyond_2023}, we treat each comment as a distinct analytical unit, categorized into one of five categories corresponding to the knowledge co-construction phases (0 = no contribution to knowledge advancement, 1–4 represent Phases 1–4). 
\mxj{
To build this classifier, we collected data from two strictly moderated communities in Reddit \cite{reddit_2005}, r/AskScience and r/ExplainLikeImFive, as they foster collaborative sense-making and iterative explanation through multi-turn dialogues, aligning well with the dynamics of knowledge co-construction.
To ensure comprehensive topic coverage, we utilized the official Reddit API to stratify our sampling across diverse domain tags inherent to these subreddits (e.g., \textit{Economics}, \textit{Social Science}), collecting a balanced number of threads per domain. We prioritized threads with high comment volume to capture the full spectrum of knowledge co-construction phases. Data collection concluded on June 25, 2025, yielding a raw dataset of 13253 comments across 225 threads. To ensure data quality, we removed comments marked as [deleted] or [removed], resulting in a refined corpus of 12587 comments.

Two authors manually coded 1,500 comments randomly sampled from the dataset.
}
They first independently labeled an initial set of 100 comments to align interpretations of phase definitions, then independently labeled the remainder. Disagreements were resolved through discussion; unresolved cases were adjudicated by a third coder. Using the 1500 labeled samples, we fine‑tuned a Mistral model \cite{jiang2023mistral} with a LoRA adapter \cite{hu2021lora}, achieving an F1-score of 82.6\% (train/test split: 70/30). This demonstrated that the classifier generalizes well across varied discussion topics.

2. \textit{Reply Identifier} — This module detects whether a new comment is connected to any earlier comment, either via explicit reply links or implicit references (i.e., addressing earlier content without using the reply function). Leveraging LLM-based semantic matching, it identifies the source comment. All comments and their identified links are stored in a dynamic graph for downstream use.

\textbf{Frequency Controller}:
This component regulates intervention timing in line with \textbf{DR3}, using two triggers:  
1. \textit{Time threshold}: if no user comments for one hour\mxj{—a threshold established in the pilot to prevent silence without dominating the conversation—}the agent intervenes to re‑activate participation and prevent prolonged inactivity.  
2. \textit{Comment threshold}: if ten new comments have been posted without the current phase reaching sufficiency, the agent intervenes; if sufficiency is reached earlier, the count resets. This mechanism addresses potential phase stagnation while avoiding unnecessary interventions when the discussion is on the right track.

\textbf{Phase-Sufficiency Evaluator}:
This module determines whether the current phase has reached sufficiency and, based on that assessment, selects the appropriate phase‑specific strategy. In our controlled lab study, the workshop‑derived sufficiency criteria for Phases 1–4 were adapted to fit the experimental context (Section \ref{sec: 4}): 

\begin{itemize}
    \item \textbf{Initiation}: 
    The goal of this phase is to elicit a diverse set of initial contributions. 
    Therefore, sufficiency is reached when more than three Phase-1 comments have occurred.
    \item \textbf{Exploration}: The desired outcome is to expand on existing ideas by developing complete argumentative structures. 
    We evaluated this based on the comment graph from \textit{Comment Analyzer}, treating each connected component of Phase-1 and Phase-2 comments as a \textit{'collective argument'} \cite{liu2023coargue}. 
    Drawing from established theories in evaluating collective argumentation \cite{weinberger_framework_2006, van2002argumentation}, each argument's structure is scored using a simplified three‑item checklist — presence of a Qualifier, Evidence, and Reasoning — with each item scored 0 or 1. A score of at least 2 marks the argument as complete. Counterarguments within a connected component are evaluated in the same way. The phase is considered sufficient when at least 70\% of collective arguments are complete. To assess LLM’s reliability in applying these criteria, two researchers independently scored Reddit comment trees, resolved disagreements, and compared the consolidated labels with the LLM’s predictions. An F1-score of 86.0\% indicated that LLM can reliably evaluate argument completeness in this context.
    \item \textbf{Negotiation}: 
    The sufficiency criterion is that every identified conflict is addressed by corresponding Phase‑3 comments
    and resolved to consensus. 
    \mxj{Specifically, the agent categorizes consensus into three types beyond simple agreement: (1) \textit{Clarified Disagreement} (where divergence is clear but irreconcilable), (2) \textit{Conditional Agreement} (where different positions hold under specific conditions), and (3) \textit{Full Agreement}.}
    LLM has demonstrated capabilities in consensus detection \cite{heller2025finding}, and we conducted a technical evaluation similar to that used for Phase 2 to verify the performance, where an F1-score of 76.7\% demonstrated its effectiveness in detecting whether a conflict has been resolved to consensus.
    \item \textbf{Co‑construction}: 
    Here, sufficiency requires that Phase-4 comments collectively summarize all agreements and contain more than three metacognitive statements that either reflect on or apply the collective knowledge.
\end{itemize}

\textbf{Style Manager}: 
Each discussion thread in the experiment is assigned a predefined intervention style (see Section \ref{sec: 4}). This module ensures that generated interventions match the assigned style.

\textbf{Response Generator}: 
When \textit{Frequency Controller} triggers an intervention, the \textit{Phase-Sufficiency Evaluator} and the \textit{Style Manager} determine which phase‑specific strategy to apply and in which style. This module then retrieves the corresponding strategy prompt from a curated set, which were developed based on workshop-derived intervention strategies (Table \ref{tab:workshop}) and refined through iterative prompt‑engineering. The selected prompt
is passed to the LLM to generate the final response.

\begin{figure*}[htbp]
  \centering
  \includegraphics[width= 0.85\linewidth]{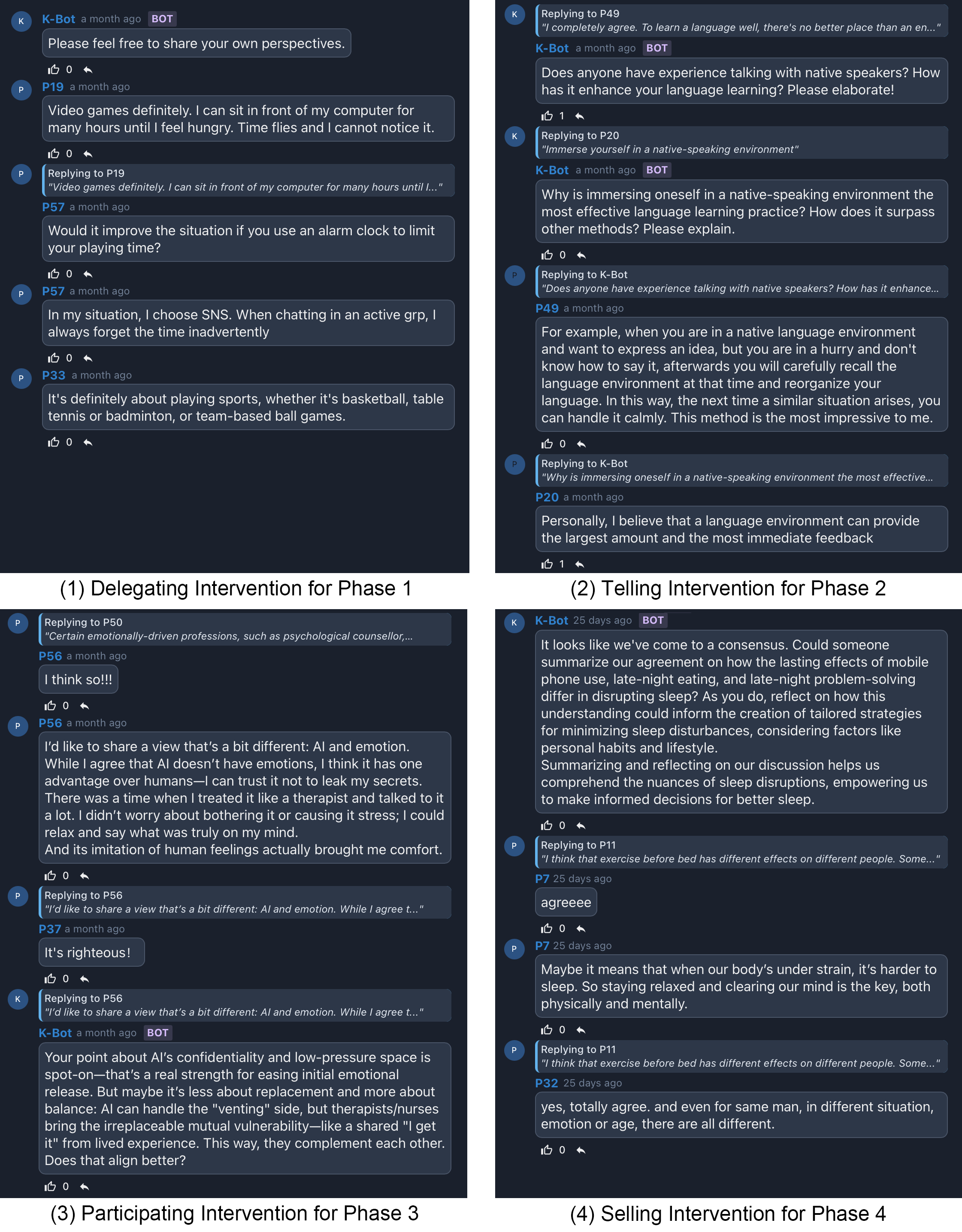}
  \caption{Example screenshots from four threads on the discussion platform used in the lab study, illustrating the interface and instances of agent interventions across different phases and styles. Participants engaged in discussions by posting comments and liking or replying to others. The examples show: (1) a delegating-style intervention tailored to Phase~1, (2) a telling-style intervention tailored to Phase~2, (3) a participating-style intervention tailored to Phase~3, and (4) a selling-style intervention tailored to Phase~4.}
\label{fig:frontend}
\end{figure*}

Overall, this architecture, integrated with the intervention logic, enables the agent to satisfy the DRs and foster knowledge co‑\allowbreak construction in asynchronous discussions.


\subsection{Experimental Platform}

\mxj{We developed a web-based experimental discussion platform with the agent integrated, accessible via standard browsers (Figure \ref{fig:frontend}), enabling participants to engage in simulated asynchronous discussions while receiving agent interventions.} In this paper, a discussion \textit{thread} denotes the entire unit comprising the initial topic post and all subsequent comments; a \textit{comment} denotes a single user message within a thread.

\mxj{Upon logging in, participants navigated from a dashboard of active threads to expanded views of specific discussions. }
The discussion interface was designed with reference to Discourse \cite{discourse}, displaying comments within a thread in strict chronological order. Compared with layouts that support nested, tree-structured replies, this structure tends to reduce interactional coherence and result in harder‑to‑follow discussions \cite{suthers2006technology, herring1999interactional}, which can hinder the progression of knowledge co-construction, providing a suitable context for the agent to exert its influence. 
\mxj{Participants could contribute new comments to the discussion via an input box at the bottom of the page or interact with specific comments using “reply” and “like” buttons. 
To nudge engagement, a notification badge in the top-right navigation bar appeared whenever a participant received a reply. Clicking this notification automatically redirected the user to the relevant comment context. }
Agent-generated comments were explicitly marked with a \texttt{BOT} tag to clearly indicate their automated origin.


\section{Experiment Design}
\label{sec: 4}

This section details our experimental setup, including conditions, study procedures, and evaluation metrics. We employed a mixed-methods, within-subject design where participants engaged in multiple consecutive discussion threads under different conditions.

\begin{figure*}[h]
  \centering
  \includegraphics[width=\linewidth]{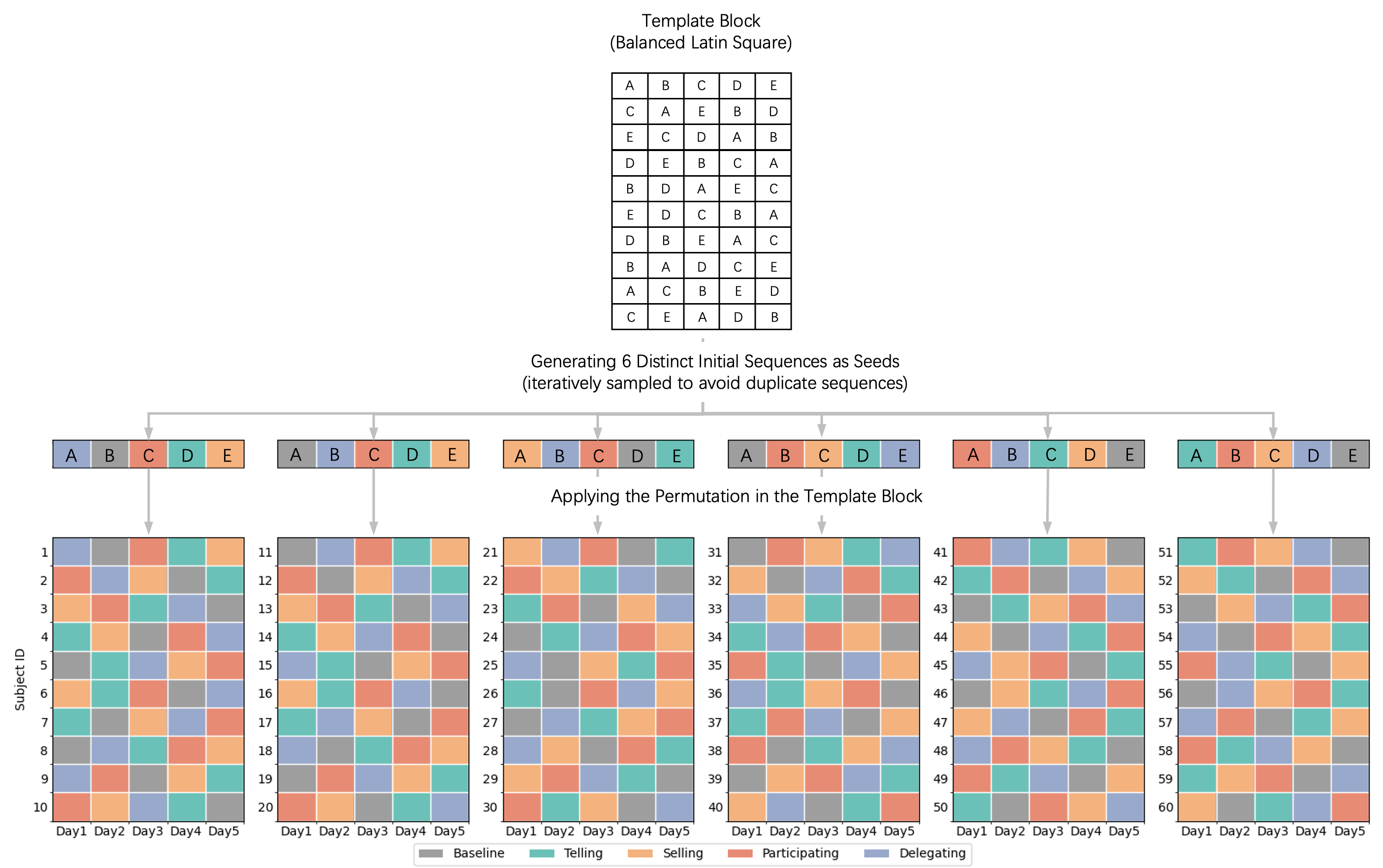}
  \caption{\mxj{The counterbalancing strategy employed in the study. We iteratively sampled initial sequences and applied the template permutation, discarding duplicates until we obtained six seeds that yielded 60 unique condition sequences.}}
  \label{fig:counterbalance}
\end{figure*}

\subsection{Conditions and Setup}
\label{sec: 4.1}
Participants experienced five conditions (four intervention styles and one human-only baseline), with each corresponding to one discussion thread. 
Studies on online communities \cite{wikipedia1percent, nonnecke2001lurkers} indicate that a small fraction of members contribute most of the content, while the majority remain passive (“lurkers”) or post infrequently. Thus, we assigned only a small number of active contributors to each thread to simulate the discussion dynamics observed in real-world forums. Pilot study showed that groups of six participants per thread yielded sufficient interaction for meaningful discourse. 

In total, we collected 50 discussion threads covering different topics. Each thread is discussed by a distinct group of 6 participants. 
No participant engaged in multiple threads simultaneously to avoid cross-thread confounding effects. Thus, the discussions are conducted sequentially. To simulate asynchronous discussions, participants were given a time window rather than a single-session requirement. Based on pilot trials, we set this duration to one day per thread. Each day, participants could log in at any time to contribute to one thread. At the end of the day, the thread was closed, and participants proceeded to a new topic under the next condition.

We selected five discussion topics, adhering to two criteria: (1) open-endedness, to naturally elicit diverse ideas and encourage negotiation; and (2) accessibility, to ensure that topics require no specialized knowledge, enabling participants to contribute meaningfully as active contributors. An example topic is “What's the most effective way to learn a new language”. A full list of topics is provided in the supplementary material.

\subsection{Participants and Procedure}

\begin{figure*}[h]
  \centering
  \includegraphics[width=0.8\linewidth]{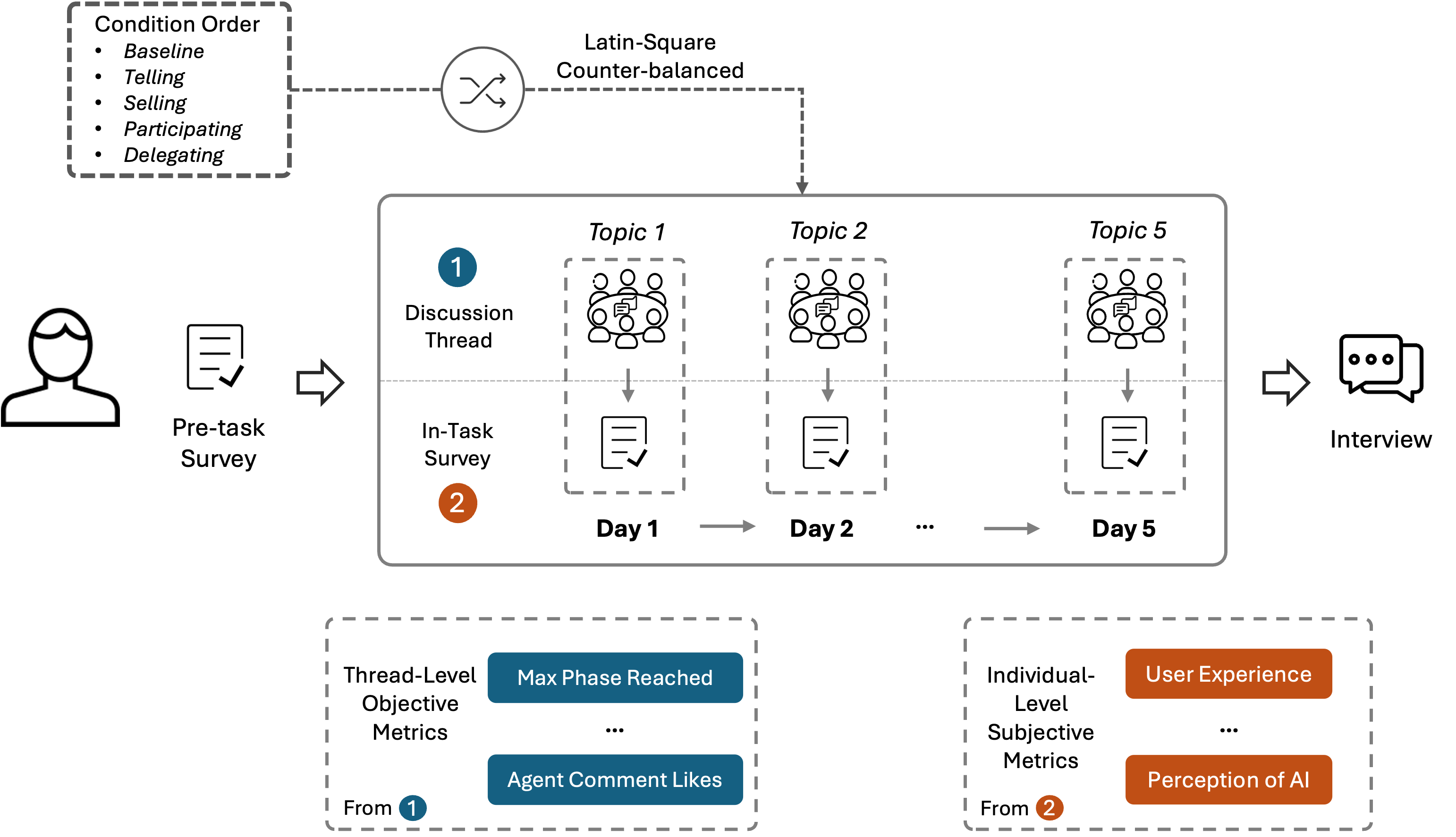}
  \caption{A within-subject study with 5 conditions (baseline + 4 intervention styles). We used a counterbalanced design to avoid order effect. Participants completed discussion threads for the assigned condition, as well as a pre-task survey, in-task surveys, and a follow-up interview.}\label{fig:study-procedure}
\end{figure*}

\mxj{We first conducted an a priori power analysis in G*Power \cite{faul2007g} based on a repeated-measures ANOVA design with five conditions. Assuming a medium effect size $f = 0.25$, a significance threshold $\alpha = 0.05$, and a statistical power $1-\beta = 0.95$, the analysis suggested a minimum sample size of 31. }
With IRB approval, we recruited 60 participants (24 female, 36 male, aged 19-33, with backgrounds in engineering, science, arts and business) through online advertisements, social media, and word-of-mouth. Participants were eligible for the study if they were active contributors of asynchronous discussion platforms. We excluded participants from our previous study to avoid potential bias.

Figure \ref{fig:study-procedure} illustrates the experimental procedure. Participants first completed a pre-task survey to collect demographic information and confirm eligibility, followed by a task briefing. They then participated in five sequential one-day discussion threads, each involving a distinct topic. 
\mxj{The topic for each day was fixed across the cohort to ensure a shared discussion context, enabling participants to interact within the same thread, with the topics being broadly accessible and of comparable difficulty (Section \ref{sec: 4.1}).
To control for order and carryover effects, we employed a \textit{balanced latin square} design \cite{williams1949experimental}. Given five conditions, a \textit{balanced latin square} block requires 10 sequences to ensure that  every condition appears in every position and precedes/follows every other condition equally often. 
To maximize interaction diversity and ensure participants encountered different peers everyday, 
we generated six mutually disjoint sets of these blocks. 
Specifically, we selected six distinct initial sequences as seeds and applied the \textit{balanced latin square} permutation to each, such that there were no duplicates among the final 60 condition sequences—one for each participant (Figure \ref{fig:counterbalance}). 
This design ensures that: (1) every condition appeared equally often on each topic, effectively neutralizing topic-specific biases; and (2) no two participants followed the exact same trajectory, preventing fixed subgroups.
}

Following each daily discussion, participants completed an in-task survey to provide condition-specific feedback. Upon completing all threads, participants took part in a semi-structured interview to explore their knowledge co-construction experiences, perceptions of different styles, and underlying rationales behind their behaviors. 
Participants were compensated at the local standard hourly rate.


\subsection{Measurements}

We evaluated each RQ using thread‑level objective metrics and individual‑level subjective metrics. All in-task surveys used a 7-point Likert scale; specific items are provided in the supplementary material.

\subsubsection{RQ1: \textit{How do AI intervention styles influence knowledge co-construction progression?}}
At the thread level, \textit{Max Phase Reached} measured the highest phase achieved, \mxj{serving as an indicator of the agent's capacity to facilitate progression rather than a requirement that every thread must attain the final phase.} \textit{Sufficiency Criteria} indicated whether the thread met predefined requirements for each phase (Table \ref{tab:workshop}). Two researchers independently coded these metrics based on the definitions in Section \ref{sec: 3}, and resolved discrepancies through discussion.
At the individual level, participants rated the discussion’s \textit{Smoothness} \cite{avula2022effects}, \textit{Effectiveness} \cite{kim2021moderator, kim2020bot}, \textit{Depth} \cite{lucas_assessing_2014}, and \textit{Sufficiency} \cite{onrubia_strategies_2009}.

\subsubsection{RQ2: \textit{How do intervention styles influence participant perceptions of the AI agent and their user experience?}}


Thread-level indicators included the total \textit{number of likes received by the agent’s comments} and the \textit{number of human replies to the agent}.
Subjective measures captured participants’ perceptions of the agent's \textit{Helpfulness} \cite{buccinca2020proxy, ma2022glancee}, \textit{Appropriateness} \cite{liu2022exploring, ma2023modeling}, \textit{Intrusiveness} \cite{do2022should}, and \textit{Social Presence} \cite{benke2020chatbot}. 
User experience was assessed through NASA Task Load Index \cite{hart2006nasa, zheng2024disciplink}, specifically \textit{Mental Demand \cite{ma2025towards}, Effort \cite{ma2025dbox},} and \textit{Frustration} \cite{ma2022glancee} following existing research, as well as \textit{Perceived Complexity} \cite{ma2024you, ma2023should} and \textit{Agency} \cite{hwang2022ai}.

\subsubsection{RQ3: \textit{How do intervention styles affect human-human interaction?}}

The primary thread-level metric was the \textit{number of human replies to other human participants}. 
Perceptions of interpersonal connection were captured through the Inclusion of Other in the Self (\textit{IOS}) scale \cite{aron1992inclusion, huang2024sharing} and perceived \textit{Mutual Awareness} \cite{peng2019gremobot}.


\begin{figure*}[t]
    \centering
    \includegraphics[width=0.45\textwidth]{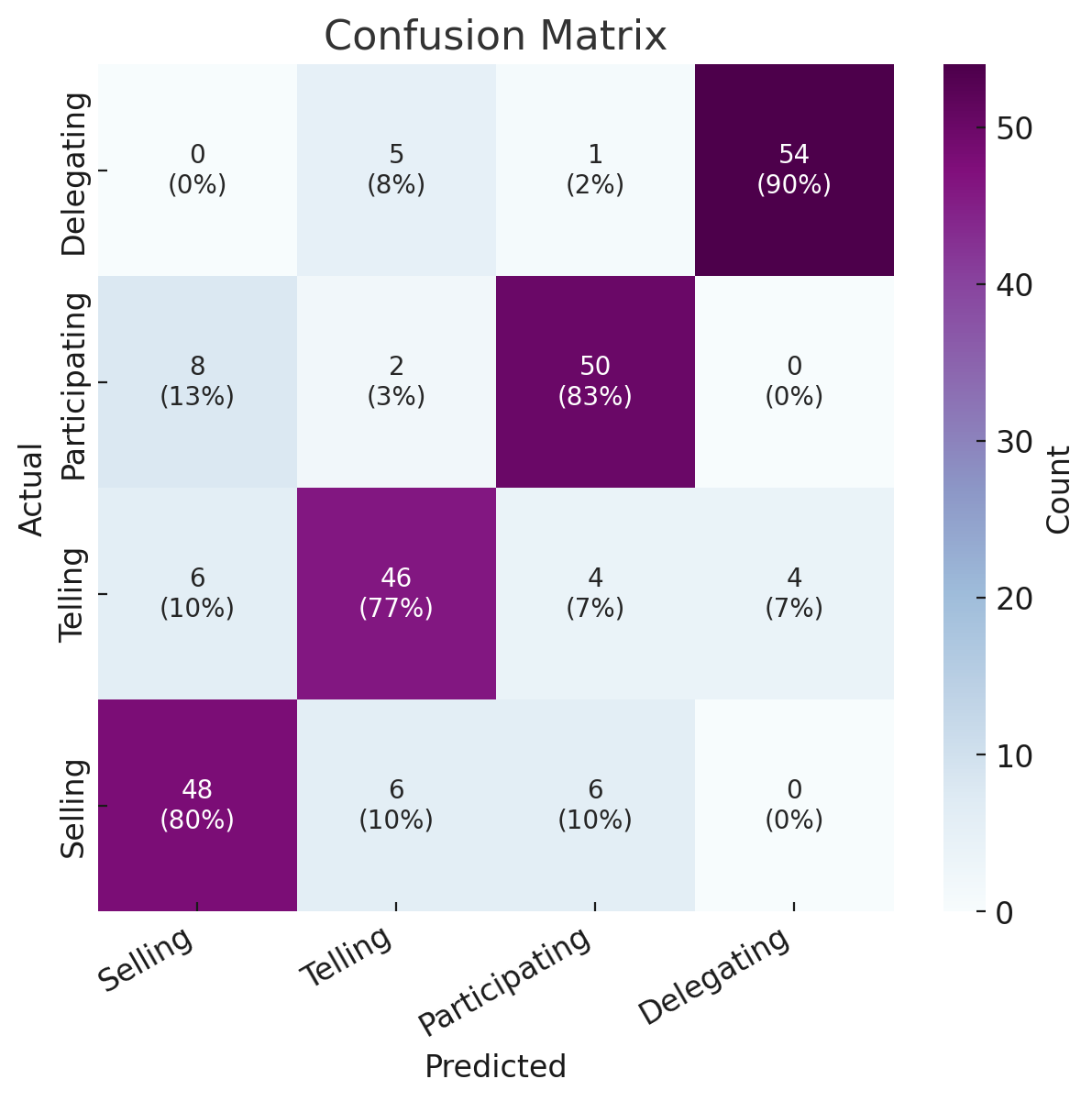}
    \caption{Confusion matrix comparing participants’ perceived intervention style of each agent with the actual style implemented. Darker cells indicate higher classification counts.}
    \label{fig:confusion_matrix}
\end{figure*}

\subsection{Analysis Methods}
We employed a mixed-methods approach for data analysis. 
\mxj{Quantitatively, for individual-level measures, Shapiro–Wilk tests \cite{shapiro1965analysis} indicated violations of normality. Consequently, we applied the Friedman test \cite{friedman1940comparison}, a non-parametric alternative suitable for our within-subjects design. Significant main effects were followed by Bonferroni-corrected Wilcoxon signed-rank post-hoc tests. We report test statistics, adjusted p-values, and effect sizes.
For thread-level metrics, we employed Generalized Linear Mixed Models (GLMM) \cite{breslow1993approximate} to model non-normal distributions without transformation and to account for data dependencies. Thread metrics were specified as dependent variables, Condition as a fixed effect, and Day as a random intercept to control for day-specific variability. We specified model distributions based on variable type: binomial for binary outcomes (e.g., Phase Sufficiency), negative binomial for count variables (e.g., Agent Comment Likes), and cumulative link models for ordinal variables (e.g., Max Phase Reached). Upon observing significant main effects, we conducted pairwise comparisons using estimated marginal means with Benjamini–Hochberg (BH) correction \cite{benjamini1995controlling}. We report model estimates, standard errors (SE), and p-values, alongside effect sizes presented as Odds Ratios (OR) for binary/ordinal outcomes and Rate Ratios (RR) for count data.
Across analyses, items for RQ1 and RQ3 were compared across all five conditions, while RQ2 focused on four intervention groups.}


\mxj{
Qualitatively, we conducted a thematic analysis \cite{braun2006using} of the semi-structured interview transcripts, grounded in each RQ \cite{hadi2022gamification, hadi2022users}. 
First, audio recordings were automatically transcribed via Zoom. Two researchers reviewed all transcripts to ensure fidelity and got familiar with the data. 
Second, they performed a structural categorization by mapping the interview segments to the corresponding RQs.
Third, within these RQ-aligned categories, the two coders independently coded all the transcripts using ATLAS.ti \cite{atlasti2023}, and frequently met during the analysis to reconcile discrepancies through negotiated agreement until consensus was reached. 
After key themes emerged from the data, they were refined through iterative team discussions. The final themes and codes are summarized in Figure \ref{fig: qualitative}, which systematically maps specific insights to corresponding RQs, with detailed interpretations and representative quotes provided in Section \ref{sec: 5}.
}

\section{Results}
\label{sec: 5}

\begin{figure*}[h]
  \centering
  \includegraphics[width=\linewidth]{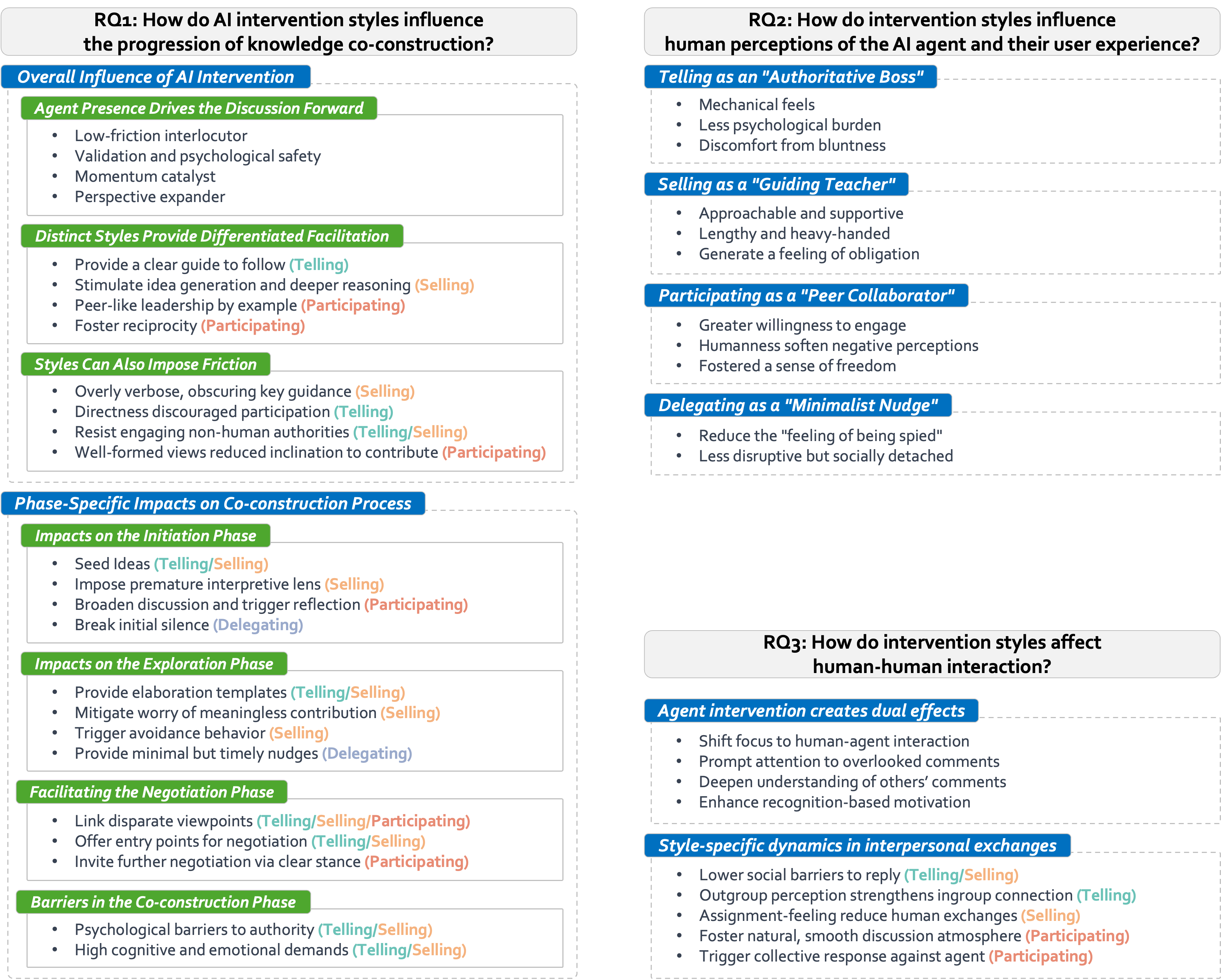}
  \caption{\mxj{Summary of the main results from the thematic analysis, organized by research questions (RQs) and reported alongside quantitative findings in the corresponding Results subsections.}}\label{fig: qualitative}
\end{figure*}

This section presents the results for each RQ, integrating statistical analyses with qualitative insights (summarized in Figure \ref{fig: qualitative}) to offer a comprehensive view. 
For brevity, we capitalize style names to refer to either agents exhibiting that style or the corresponding condition; for example, “Selling” denotes selling‑style agents or the selling condition.


\subsection{Manipulation Check}

To validate our manipulation, we compared participants’ judgments of each agent’s intervention style with the style that the agent was programmed to enact, as shown in Figure~\ref{fig:confusion_matrix}. Overall, 82.5\% of classifications matched the intended style, suggesting that participants generally recognize the correct style. A Chi-square test confirmed that the observed classification frequencies significantly differed from chance ($\chi^{2}(9) = 432.58, p < .001$), indicating that the agents were reliably perceived in line with their designated styles. 
This verified that our manipulation of the independent variable was effective, ensuring that our participants perceived, understood, and reacted to the four intended agent styles as expected.

\subsection{RQ1: How do AI intervention styles influence the progression of knowledge co-construction?}


We first compared discussions with AI intervention (four styles) against the no-agent baseline. We then examined differences among the styles in shaping the overall knowledge co-construction process, followed by a phase-level analysis of how each style facilitated progression. 

\begin{table*}[ht]
\centering
\includegraphics[width=0.83\linewidth]{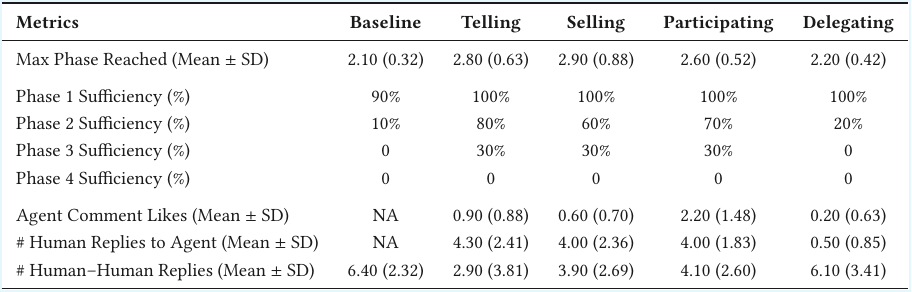}
\caption{Descriptive statistics of thread-level measures across the five experimental conditions. The metrics \textit{Agent Comment Likes} and \textit{Number of Human Replies to Agent} are reported only for the four intervention-style conditions, as the baseline condition does not include an agent.}
\label{tab:descriptive}
\end{table*}

\begin{table*}[ht]
\centering
\includegraphics[width=0.58\linewidth]{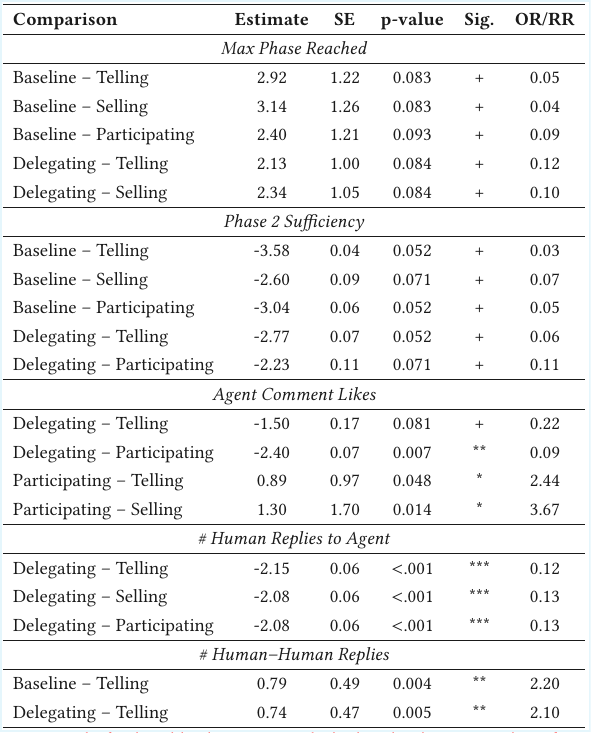}
\caption{\mxj{Pairwise comparison results for thread-level measures. \textit{Only dyads with at least marginal significance (p~$<$.10) are included.} Reported values include the model estimate, standard error (SE), p-values (+: .050~$<$~p~$<$.100, *: p~$<$.050, **: p~$<$.010, ***: p~$<$.001) adjusted using the Benjamini–Hochberg (BH) correction procedure,, and effect sizes (odds ratios for ordinal/binary variables and rate ratios for count variables).}}
\label{tab:pairwise}
\end{table*}

\subsubsection{Discussion Dynamics with Agent Intervention}

\begin{figure*}[h]
  \centering
  \includegraphics[width=\linewidth]{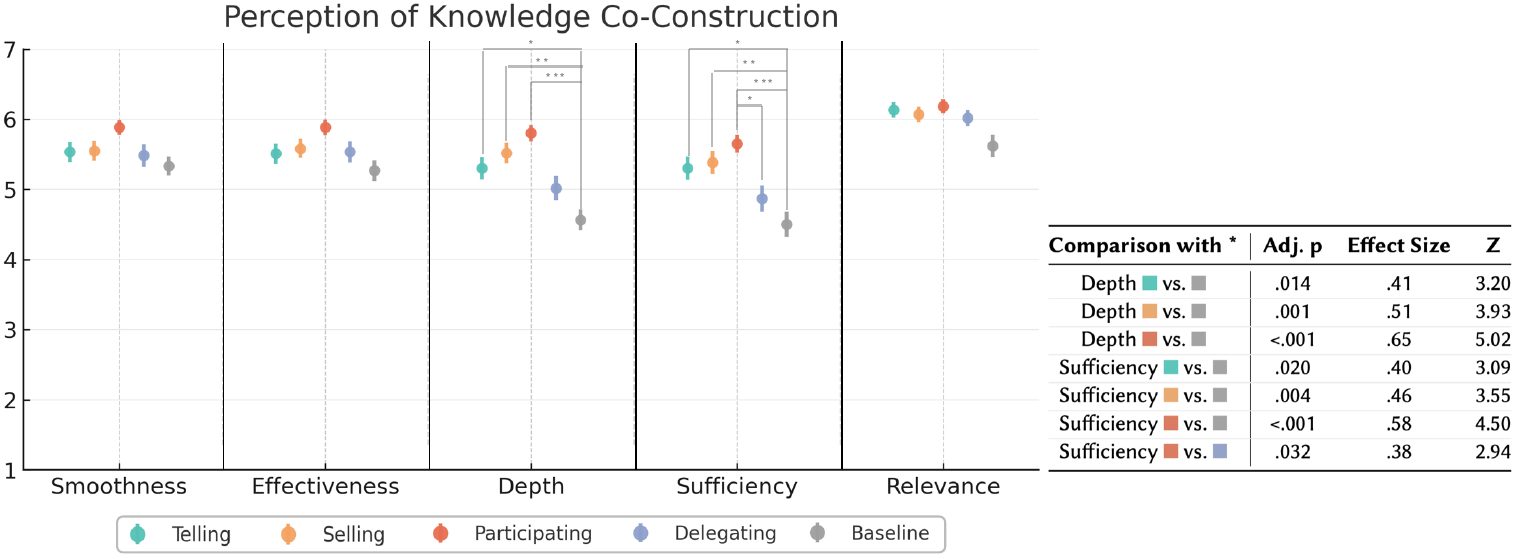}
  \caption{\mxj{Participants’ perceptions of knowledge co-construction across conditions. Error bars indicate 95\% confidence intervals (*$p<.050$, **$p<.010$, ***$p<.001$). For significant pairwise comparisons, test statistics and effect sizes are reported. For all items, a higher score indicates better performance.}}
  \label{fig:K_perception}
  \label{fig:9}
\end{figure*}

Table \ref{tab:descriptive} presents descriptive statistics of the thread-level measures, and Table \ref{tab:pairwise} reports pairwise comparisons (only dyads with at least marginal significance are shown for brevity; full results are provided in the supplementary material). 
As shown in these tables, all intervention conditions other than Delegating was associated with markedly deeper progression of knowledge co-construction than the baseline. The average \textit{Maximum Phase Reached} increased by 23.8–38.1\% (Table \ref{tab:descriptive}); pairwise comparisons reached marginal significance with 
\mxj{substantial odds ratios (ORs: 0.04 – 0.09) (Table \ref{tab:pairwise})}, suggesting practically meaningful improvements despite not meeting conventional thresholds. Baseline discussion typically plateaued in the early stages, with only rare comments reaching Phase 3 or beyond, whereas agent-intervened threads consistently advanced further through the phases.
Phase-sufficiency analysis reinforces this divergence. While 90\% of baseline groups met sufficiency in Phase 1, only 10\% reached sufficiency in Phase 2. In comparison, Telling, Selling, and Participating conditions yielded \textit{Phase 2 sufficiency} in 60–80\% of cases,
\mxj{reaching marginal significance against baseline yet showing substantial differences in odds (ORs: 0.03–0.07).} (Table \ref{tab:pairwise}). Moreover, these three styles uniquely achieved Phase 3 sufficiency in about 30\% of threads, an outcome never observed without an agent. This indicates that agents not only increased the likelihood of advancing deeper but also strengthened earlier phases as foundations for further progress.
These findings align with the subjective evaluations (Figure \ref{fig:K_perception}). Compared with the baseline, participants perceived significantly higher levels of \textit{Depth} and \textit{Sufficiency} of knowledge co-construction under Telling, Selling, and Participating interventions, while Delegating showed no significant difference.

Interview feedback revealed why the agent was able to advance knowledge co-construction. First, as a \textbf{low-friction interlocutor}, the agent lowered the social cost of speaking up, making it easier to post the \textit{“first sentence”} (P37, P42, P59). As one participant noted, \textit{“When discussing with AI, I didn’t feel any burden or that I was inconveniencing others”} (P37).
This consequently expanded the pool of visible contributions at the Initiation stage, laying the groundwork for subsequent enrichment and deepening of knowledge.
Once participants began contributing, the agent provided \textbf{validation and psychological safety}. Even brief replies by the agent conveyed that those contributions were seen and valued, alleviating the perceived risk of being ignored and encouraging participants to elaborate. \textit{“When the bot replied, I felt more at ease… even a bot’s response made me feel my comment was acknowledged”} (P17). 
Another added, \textit{“Receiving [bot's] reply… prompted me to explore the topic more deeply” }(P40).
Such reassurance not only offered emotional comfort but also made it seemingly worthy of expanding tentative ideas into more detailed and substantive contributions.
When discussions slowed down, the agent acted as a \textbf{momentum catalyst} that reactivated stalled threads, provided footholds for others to pick up and continue, and kept ideas flowing. As P42 observed, \textit{“When progress was slow, AI motivated us and kept the discussion going”}. Others emphasized that without such prompts, \textit{“many continuations wouldn’t have happened”} (P12). 
Beyond maintaining rhythm, the agent also functioned as a \textbf{perspective expander}. It surfaced overlooked angles and concrete details (P12). Engagement was strongest when prompts resonated with existing thinking—\textit{“I respond when the bot’s reply connects with my cognition”} (P17). 
The agent also introduced alternative viewpoints, nudging participants away from one-sided statements and toward considering multiple perspectives. \textit{“Humans don’t usually present both pros and cons; the bot took that role and guided me to consider the opposite view”} (P4). 
Meanwhile, not all participants embraced the agent as a true member of the group, as P56 put it, \textit{“I wouldn’t reply to a bot; I’d rather talk to people”}. This reluctance varied among different agent styles, a contrast we present further in Section \ref{sec:6.2.2}.

\subsubsection{Differential Impacts of Intervention Styles}
\label{sec:6.2.2}

Having examined the effect of agent presence in general, this section compares how the four styles differentially shaped knowledge co-construction.
\mxj{Overall, both task-oriented styles—Telling and Selling—marginally outperformed Delegating in \textit{Max Phase reached}, and Telling retained this advantage for \textit{Phase 2 Sufficiency} (Table \ref{tab:pairwise}). 
Despite these statistical similarities, participants perceived the two styles differently. 
Selling was appreciated by some for its elaborative explanations, which they felt \textbf{stimulated idea generation and deeper reasoning}. As P23 explained, \textit{“These explanations broadened my thinking and made it easier to follow the agent’s reasoning.”} However, others found it \textbf{overly verbose, obscuring key guidance} (P31, P52). 
Telling, in contrast, was praised for its clarity and conciseness. As P60 noted, \textit{"\textbf{Providing a clear guide to follow} makes the process much easier”}. Yet, this style also had drawbacks. }
A few participants questioned its impact, noting that its brevity was easy to ignore, and its \textbf{directness discouraged participation} (P17-19, P56). As P17 explained, \textit{“It urged me like a task to finish, which suppressed my willingness to contribute”}."
Notably, a group of participants (13/60) explicitly \textbf{resisted engaging} with either style, perceiving them as \textit{“non-human authorities"}. \textit{“I don’t want to spend energy on an AI”}, said P59.

\mxj{Compared with Delegating, Participating approached a significant improvement in \textit{Phase 2 Sufficiency} (Table \ref{tab:pairwise}) and was perceived as fostering greater \textit{Sufficiency} in discussion (Figure \ref{fig:K_perception}).} Despite its low task-orientation, it achieved outcomes comparable to more directive styles. Many participants (37/60) attributed this to its resemblance to a typical human member, often treating the agent as a peer. Some even reported forgetting that it was a bot, \textit{“It sounded so much like a person that I just replied instinctively”} (P26). This peer-like quality encouraged \textbf{leadership by example} (P10, P23) and created \textbf{an equal, immersive interaction that fostered reciprocity}, generating a virtuous cycle of contributions (P8, P26, P53).
However, some cautioned that its guidance was too subtle, risking stagnation when stronger direction was needed (P39, P42). Others noted that when the agent presented well-formed viewpoints, it paradoxically \textbf{reduced their inclination to contribute further} (P21, P59). As P21 noted, \textit{“It felt so smart that I didn’t need to add anything”}.

Delegating showed no significant differences from the baseline across all metrics (Table \ref{tab:pairwise}), indicating minimal impact on discussion progression. Most participants (44/60) reported largely ignoring its interventions due to minimal presence, as P37 remarked, \textit{“It didn’t affect our discussion”}.

\subsubsection{Phase-Specific Effects of Intervention Styles}
\label{sec:6.2.3}
Beyond overall effects, we examined how each style influenced specific phases of knowledge co‑construction through qualitative insights.
\paragraph{Initiation} 
In this phase, participants introduce perspectives that set the stage for subsequent discussion. 
Telling and Selling often functioned as \textbf{idea seeders}, lowering the entry cost of participation and diversifying discussion starting points. As one participant explained, \textit{“It gave me a seed on something unfamiliar but interesting. I looked it up, reflected, and shared on the forum; without the bot’s comment, I wouldn’t have done so”} (P17). However, Selling’s additional explanations sometimes imposed a \textbf{premature interpretive lens}, which participants felt could constrain the openness of thought and channel ideas into narrower tracks (P36). By contrast, Telling was praised for offering concise prompts without persuasive framing, thereby \textit{“providing direction while leaving room for user elaboration”} (P36, P53).
Participating contributed its own viewpoints to \textbf{broaden discussion} (P3, P21); when its perspective deviated from participants' existing cognition, it often \textbf{triggered deeper reflection}. \textit{“When the bot’s view was inconsistent with mine, it pushed me to think more deeply about why it was not the same. Unlike with humans, I felt no pressure to oppose it, so I could easily strengthen my own position from the opposite direction”} (P4).
Delegating mainly \textbf{broke initial silence} through a simple reminder (P40, P42), without substantive direction to shape early-stage contributions.

\begin{figure*}[tp]
  \centering
  \includegraphics[width=0.9\linewidth]{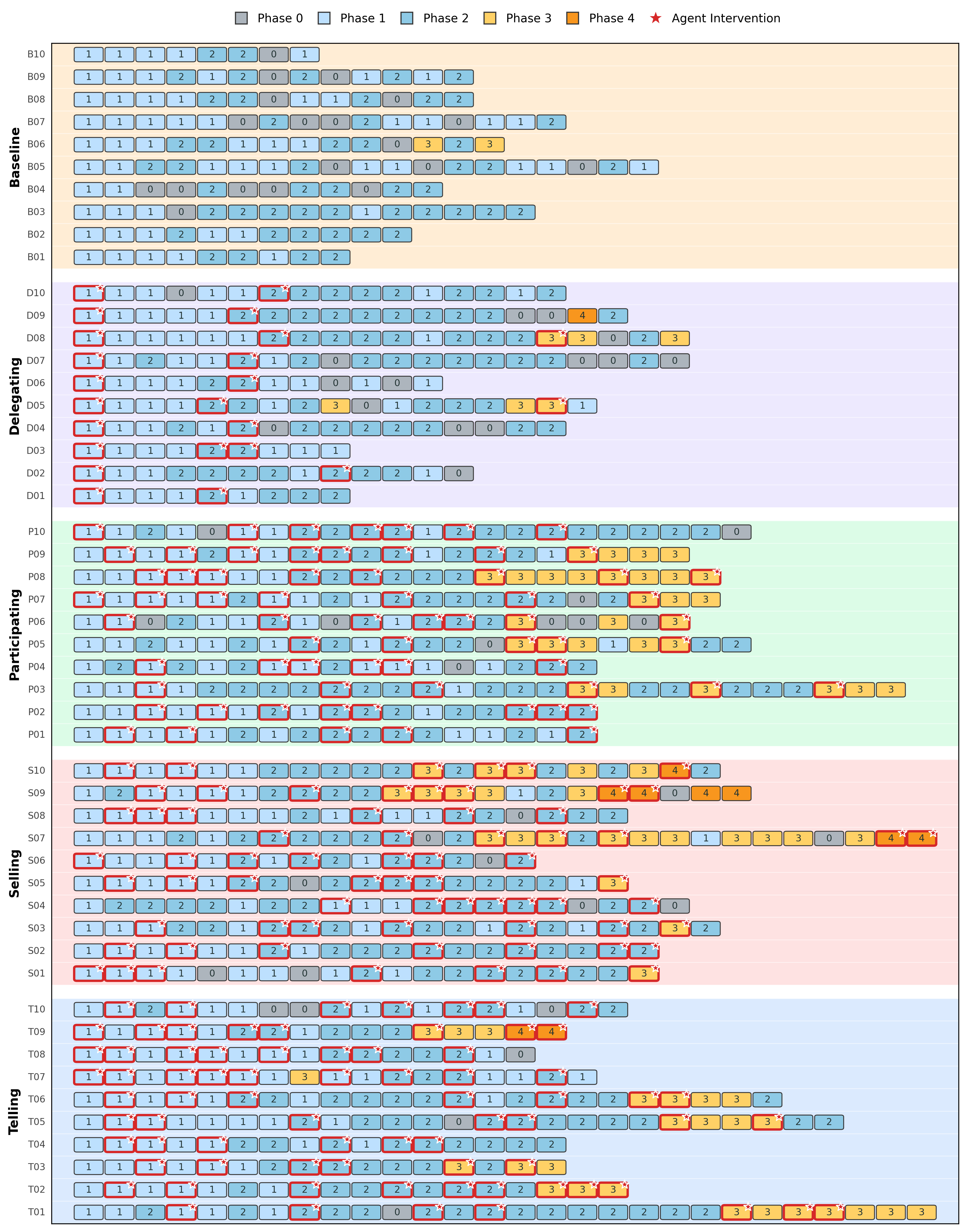}
  \caption{Chronological comment sequences for each thread across different conditions, showing the process of knowledge co-construction. Each cell represents a comment, with the number indicates the phase it is categorized into through manual coding. Red-bordered cells with a red star denote agent interventions, where the number indicates the targeted phase. Colors are also applied to encode phases.}
  \label{fig:swimline}
\end{figure*} 

\paragraph{Exploration}

In this phase, participants expanded previous contributions into more complete arguments.
Telling and Selling were often described as offering a \textbf{\textit{“template”} for enriching arguments}, triggering memories of relevant experiences or knowledge participants might not have otherwise recalled (P43).
For Selling, the added persuasion prompted more supplementary comments since it \textbf{mitigated the worry of \textit{“contributing something meaningless”}}(P20) through explicit justification of why a point merited further elaboration.
Yet the same detailed justifications could also make participants feel \textit{"compelled to respond"} as if the bot’s reasoning left little room to disengage or disagree (P3, P25, P59). 
Some participants reacted with \textbf{avoidance behavior} — skimming or ignoring the intervention, withholding replies, or posting perfunctory, low-effort responses merely to close the loop (P25, P59).
As P59 noted, \textit{“It explained so thoroughly that I felt obliged to reply, almost as if it was forcing me to accept its instructions”}. 
This avoidance might help explain Selling's lower rate of \textit{Phase 2 sufficiency} compared with Telling and Delegating (Table \ref{tab:descriptive}). One typical case was S04 (Figure \ref{fig:swimline}): despite five consecutive Phase-2 interventions by Selling, subsequent replies were either absent or superficial, with only one substantive Phase-2 comment.
By contrast, Delegating was sometimes appreciated for \textbf{providing minimal but timely nudges} that prompted transition to Phase 2 (P16, P51).
Several clear examples emerged, such as D08–10 in Figure \ref{fig:swimline}, where comments were mostly in Phase 1 before the Delegating prompt, but afterward shifted toward Phase 2.
From the interview, Participating showed no distinct effect in this phase, with contributions resembling its overall impact.

\paragraph{Negotiation} 
In this phase, participants worked toward resolving conflicts.
All three styles—Telling, Selling, and Participating—were valued for \textbf{linking disparate viewpoints} and making participants notice perspectives they might otherwise have overlooked (P2, P44). Telling and Selling provided \textbf{actionable entry points for negotiation}. 
\textit{“They gave me angles to negotiate on, making it clearer how to start and synthesize previous comments”} (P33, P42).
Participating operated differently, \textbf{carrying a clear stance on reconciling differing viewpoints}. This stance served as a concrete reference for further negotiation (P42, P58) and provoked deeper involvement—\textit{“its words were quite absolute, which made me want to negotiate with it”} (P56)—while others felt compelled to expand the discussion because \textit{“it wasn’t necessarily the most authoritative answer”} (P17). By simulating a realistic discussant, it prompted participants to \textbf{articulate counterarguments and refine their positions}. \textit{“when it challenged my stance, it naturally pushed me to respond”} (P18). This dynamic made Participating particularly effective for preventing stagnation in Phase 2 and facilitating transitions into Phase 3. As seen in Figure \ref{fig:swimline}, nearly all threads with a Phase-3 intervention by Participating received follow-up Phase-3 comments. By contrast, in threads like S09 or T02, Telling or Selling often required multiple consecutive Phase-3 interventions to elicit even a single negotiation comment, and sometimes elicited none.

\paragraph{Co-Construction}
In this phase, participants synthesize and reflect on the group’s shared knowledge. In our experiment, Phase‑4 interventions appeared only in four threads under the Telling and Selling conditions—consistent with their task‑oriented nature—and elicited just one Phase‑4 response.  
Many users viewed summarizing and reflecting as responsibilities for senior members or AI itself, rather than ordinary members (P2, P23). Summarization was seen as an evaluative act that implicitly claims authority over the group’s knowledge, creating \textbf{psychological barriers} as participants worried about overstepping or being judged on the quality of their synthesis (P42). 
Participants also noted that producing a synthesis carried substantially \textbf{higher cognitive and emotional demands} than the lighter, incremental contributions of earlier phases, further discouraging engagement (P51).
Consequently, directive behaviors that explicitly asked participants to summarize or reflect, whether through Telling’s blunt prompts or Selling’s persuasive framing, were largely ineffective in transitioning discussions into the co-construction stage.

\subsection{RQ2: How do intervention styles influence human perceptions of the AI agent and their user experience? 
}
\begin{figure*}[h]
  \centering
  \includegraphics[width=\linewidth]{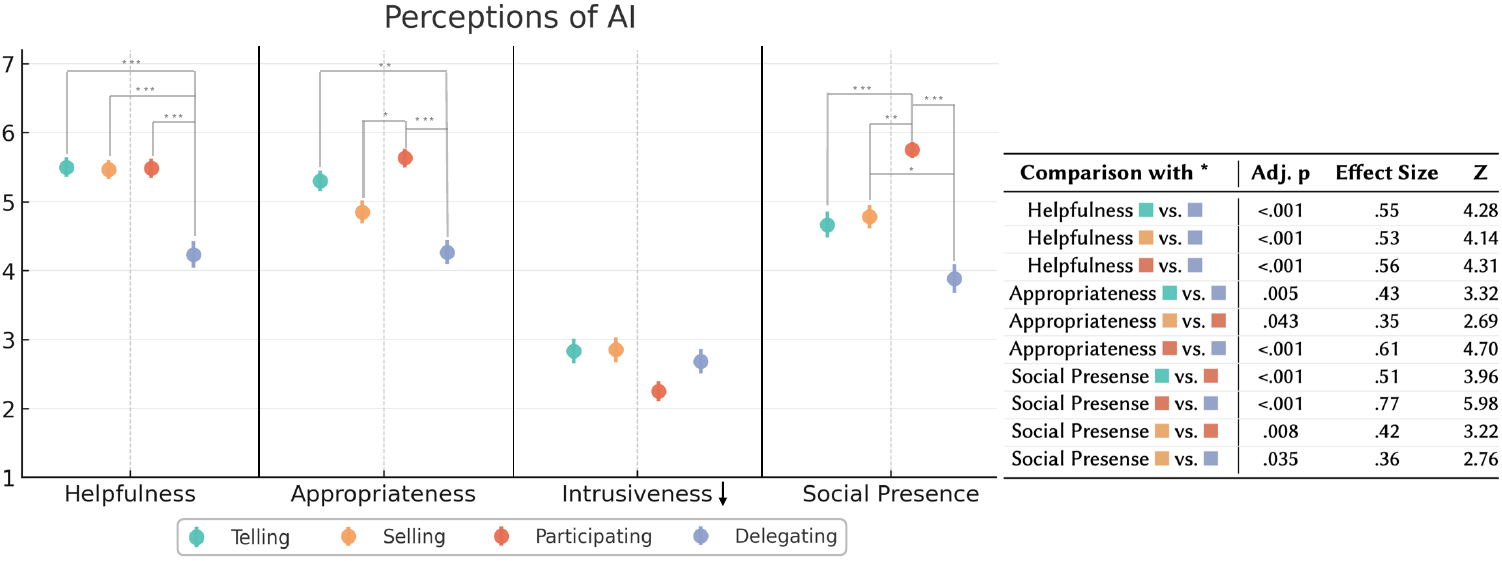}
  \caption{\mxj{Participants’ perceptions of agents across styles. Error bars represent 95\% confidence intervals (*$p<.050$, **$p<.010$, ***$p<.001$). For significant pairwise comparisons, test statistics and effect sizes are reported. Unless marked with $\downarrow$ (indicating lower is better), higher scores denote better performance.}}
  \label{fig:AI_perception}
\end{figure*}

\subsubsection{\mxj{Perception of Agent Styles}}

Participants’ perceptions of the agents varied across styles, with multiple pairs showing notable differences in both subjective and objective measures.
\mxj{Delegating elicited significantly fewer human replies compared with the other three conditions (Table \ref{tab:pairwise}). 
Subjectively, it was consistently rated as significantly less \textit{Helpful}, less \textit{Appropriate} (except vs. Selling), and lower in \textit{Social Presence} (except vs. Telling) than all other styles.}
This aligns with its minimal, mechanical, and low-frequency design. Participants often described it as \textit{“system-generated tips”} rather than a genuine member (P13, P34, P41). They acknowledged such \textbf{Minimalist Nudge} as  \textbf{less helpful or socially engaging} but also \textbf{less disruptive}; several even noted that this minimalist presence \textbf{reduced the \textit{“feeling of being spied”}} compared with other agents (P8, P21).

\mxj{Participating, by contrast, received significantly more \textit{Likes} than the other three styles (Table \ref{tab:pairwise}). It was rated more \textit{Appropriate} than Selling and higher in \textit{Social Presence} than both Telling and Selling (Figure \ref{fig:AI_perception})}. As discussed in Section \ref{sec:6.2.2}, this can be attributed to its perceived humanness. Participants tended to treat it as a \textbf{Peer Collaborator} offering equal and natural interaction. P23 noted that, \textit{“It felt like it was discussing with us rather than leading us [unlike Selling or Telling].”}. Although \textit{Human Reply Counts} did not differ significantly between Participating and Telling/Selling, interviews revealed a  \textbf{greater willingness to engage} with Participating’s comments because they resembled human dialogue (P33, P43), whereas responses to Selling/Telling often felt more like compliance with instructions (P8, P46). 
This perceived humanness also \textbf{softened aspects that were viewed negatively} in other styles. For instance, several participants complained that Selling or Telling sometimes replied selectively, which felt unresponsive —\textit{“as if the AI should address everything”} (P12). By contrast, when the Participating agent did not respond to a comment, participants interpreted it as a natural human-like choice rather than a system flaw. P49 explained that, \textit{“Even if it skipped my comment, I saw it as a natural reaction, not an AI defect.”} Similarly, while some participants (14/60) noted that Selling’s long message created cognitive load, they tolerated Participating's - \textit{“Even if its messages were long, they didn’t feel tiring to read”} (P31).

Selling and Telling did not differ significantly in quantitative measures, but participants’ perceptions diverged. Telling was often described as resembling an \textbf{Authoritative Boss}, distant, and pushy, which sometimes provoked negative emotional reactions (P17, P37). Some explicitly called its tone \textit{“cold”} or \textit{“rigid,”} highlighting its \textbf{mechanical feel} (P17, P24). Selling, by contrast, was frequently likened to a \textbf{Guiding Teacher}. Participants felt discussions with its interventions resembled classroom activities, sometimes  supportive but also at times verbose and overly didactic (P18, P26, P52). While some appreciated its \textbf{\textit{“kind and approachable”} tone} (P15), others found it \textbf{\textit{“long-winded”} and heavy-handed} (P25).

\subsubsection{User Experience}
\label{sec:UX}

As shown in Figure \ref{fig:UX}, there were no significant differences across the five conditions in perceived \textit{Effort, Frustration, Complexity}, or \textit{Agency}. \mxj{However, participants reported significantly higher \textit{Mental Demand} for Selling compared with Baseline}. As elaborated in Section \ref{sec:6.2.3}, Selling's detailed, persuasive explanations made participants feel they could not skip them, creating \textbf{a sense of obligation}—\textit{“hard to reject”}—that increased emotional burden. The \textbf{relatively long messages} further contributed to this demand.
\mxj{Regarding Telling, participants noted that its \textbf{conciseness created less psychological burden}}; its comments could be \textit{“taken or left”} and thus were not perceived as disruptive (P8). However, when Telling addressed sensitive topics, its \textbf{bluntness produced discomfort}. P39 recalled, \textit{“especially when it brought up real-life jobs, it felt too specific and uncomfortably close.”}
Participating was valued for its non‑directive style, which neither enforced agreement nor imposed constraints, fostering \textbf{a sense of freedom}: \textit{“I felt free to engage with it or not, without pressure”} (P56).

\subsection{RQ3: How do intervention styles affect human-human interaction?}

\begin{figure*}[h]
  \centering
  \includegraphics[width=0.8\linewidth]{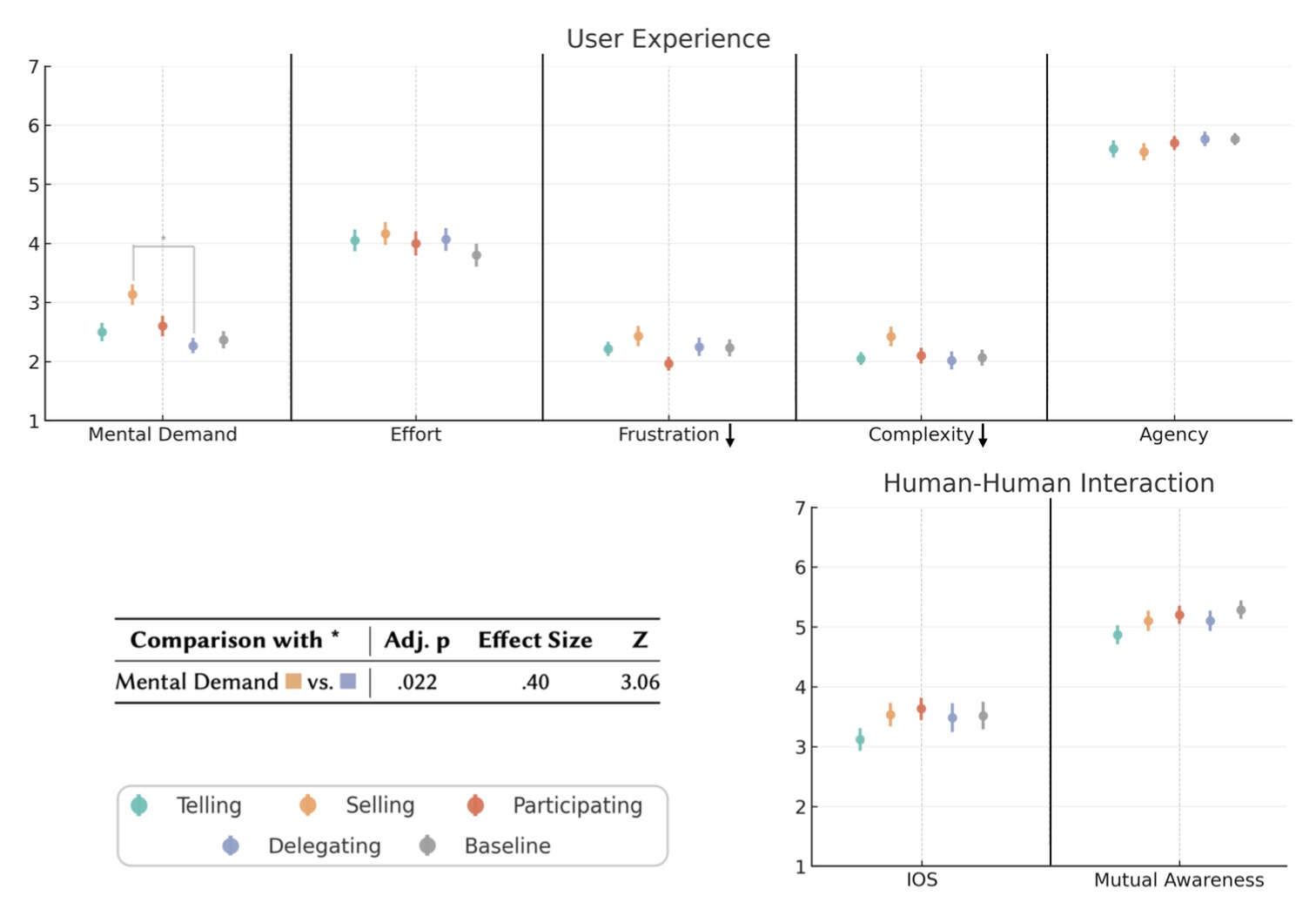}
  \caption{\mxj{Participants’ perceptions of (a) their user experience and (b) human-human interactions in the discussion across different conditions. Error bars represent 95\% confidence intervals (*$p<.050$, **$p<.010$, ***$p<.001$). For significant pairwise comparisons, test statistics and effect sizes are reported. Unless marked with $\downarrow$ (indicating lower is better), higher scores denote better performance.}}
  \label{fig:UX}
\end{figure*}
\mxj{
While subjective ratings of human–human interaction showed no significance (Figure \ref{fig:UX}), thread-level data revealed that Telling elicited significantly fewer human–human replies compared with both Baseline and Delegating (Table \ref{tab:pairwise}).
This result resonates with a broader concern raised in interviews.
Some participants (17/60) mentioned that the agent’s interventions—particularly when manifesting as concise and clear comments like Telling's—\textbf{shifted attention toward human–agent rather than human–human exchanges}.
As P60 noted,
\textit{“People chose to interact with the bot, distracting from interactions with humans”}.
When the AI replied to a human comment, subsequent contributions often addressed the AI instead of the original commenter (P21, P40). In such cases, the AI tended to substitute for, rather than mediate, peer interaction, leaving participants with fewer opportunities to engage directly with one another (P54).
Yet, this disruptive effect was not absolute. In many other instances, users viewed agents as stimulants for interaction. 
Several participants felt that agents activated discussion and created more opportunities for mutual replies (P18–20, P59). 
}Moreover, when agents (except Delegating) replied to human comments, participants perceived higher quantity and quality of peer exchanges, for two main reasons: (1) such replies  \textbf{drew attention to comments} that might otherwise have been overlooked, prompting others to engage (P30, P31, P41); and (2) by elaborating on prior comments, the agent \textbf{helped others better understand them}, leading to more substantive responses. As P46 explained, \textit{“At first some comments were short and unclear, but after the AI’s reply I understood them and wanted to respond seriously.”}
Some also noted that the \textbf{recognition signaled by bot replies} also encouraged more idea exchange (P7, P60). As P60 put it, \textit{“When it replied to me, I felt acknowledged, and became more willing to interact with others.”}


Participants also highlighted style-specific insights. For Phase-2 interventions of Telling and Selling, directives or persuasion targeted at specific human comments \textbf{lowered the barrier for some to reply} to those messages (P3–5, P17, P33). As P17 explained, \textit{“I usually feel awkward supplementing others’ comments because I fear criticism, but with the bot’s directive it felt natural”}. 
In such cases, participants attributed their decisions to interact with others to the bot, which implicitly reduced their perceived risk of being criticized (P5, P33). 
Telling sometimes reinforced human–human exchanges by being perceived as more detached from the human discussants, subtly \textbf{strengthening an \textit{“ingroup–outgroup”} dynamic} \cite{tajfel2001integrative}. As P24 put it, \textit{“The AI was on the opposite side, so we paid more attention to our human‑to‑human communication.”}
By contrast, Selling \textit{"resembled a teacher initiating classroom activities"} (P22);  participants \textbf{treated its interventions as assignments to complete}, rather than opportunities for peer dialogue, which consequently diminished human–human exchange.
Participating was perceived as fostering \textbf{more \textit{"natural"} and \textit{"smoother"} communication atmosphere} (P53). Moreover, the agent's own viewpoints occasionally \textbf{triggered collective response}—opposing or debating its views—which some felt strengthened group cohesion: \textit{“When we all refuted it together, our connection felt closer”} (P37).
Delegating was appreciated as a \textit{“hands-off”} style that left interactions undisturbed. However, participants noted interactions often remained shallow in this condition, consisting mainly of repetition or simple agreement (P8, P39).


\section{Discussion}
Our evaluation demonstrates that LLM agents equipped with phase-sensitive strategies can reliably advance asynchronous discussions toward deeper phases of knowledge co-construction. Task- and relationship-oriented styles each exhibit distinct strengths and limitations. Drawing on these findings, we first highlight the effectiveness of the proposed process-orchestrated design, then discuss its limitations and generalizability. We next examine the tensions between different styles and explore potential resolutions, and finally propose key design considerations for future research.

\subsection{Process-Orchestrated, Phase-Sensitive Agent Design}

The agents adopted a simple but effective intervention logic: \textbf{ensure sufficiency in the current phase before explicitly pivoting to the next phase}.
This differs from prior work that optimizes for single-phase behaviors. For example,the agent proposed by Ito et.al. \cite{ito2022agent} prompts for elaboration and clarification within the exploration phase; however, without an process-oriented orchestration, continued use of such cues after arguments are well-formed may restrain the natural progression toward integrative phases.
Our experiment verifies that such process-orchestrated design can deepen knowledge co-construction in asynchronous settings. Participants recognized the agent’s intent to advance the discussion, even through Delegating interventions, which often acted as \textit{"timely situational cues"} (P16). Several noted that phase transitions felt natural because preceding phases were sufficiently developed (P23, P57), enabling smooth, organic progression of discussion threads.
This orchestration logic is especially pertinent in asynchronous contexts. Unlike synchronous discussions, where shared temporal context can naturally drive progression \cite{fuks2005applying}, asynchronous platforms allow participants to join a thread at various stages, which may be underdeveloped, requiring elaboration, or already well-formed and ready for integration. At the collective level, such asynchronous entry points can lead to misaligned focus across participants—for example, late joiners to an advanced thread may revisit foundational issues, while long-time contributors are already synthesizing ideas. This divergence can fragment the discussion into parallel, uncoordinated sub-conversations, reducing the efficiency of collective knowledge advancement.
By making phase readiness and transitions explicit, the agent maintains shared orientation, supporting steadier progression toward deeper co-construction.


However, our implementation revealed a limitation: mismatches between the agent’s sufficiency judgments and participants’ perceived readiness. Sometimes when discussions stagnated in Phase 2, the agent triggered a Phase-2-to-Phase-3 transition, some participants experienced it as interruptive because they had not yet \textit{“talked it through”}. To mitigate such dissonance, future designs could \textbf{humanize sufficiency assessment by incorporating social readiness signals}—such as participation dispersion, recent uptake of new ideas, or decreasing redundancy—alongside content-based indicators. Such adjustments may preserve perceived autonomy and reduce psychological reactance \cite{deci2000and, brehm1966theory}, while still enabling timely, phase-appropriate progression.


Although developed for knowledge co-construction, the process-orchestrated logic can extend to other multi-phase collaborative tasks where later stages presuppose the successful completion of earlier ones. Examples include collaborative writing 
\cite{lowry2004building} and project-based online learning \cite{blumenfeld1991motivating}. 
However, tasks without sequential dependencies—such as parallel idea listing or casual social chatting—are less suitable. In these cases, non-linear exploration is a core affordance \cite{scott2011sage}, and imposing staged progression may constrain creativity and disrupt flow \cite{csikszentmihalyi1997flow, bederson2004interfaces}. Similarly, in time-critical collaboration (e.g., emergency response coordination \cite{turoff2004design}), our intervention logic may be counterproductive, as speed of convergence outweighs the value of building exhaustive grounding.

\subsection{Navigating the Tension between Task-Oriented and Relationship-Oriented Intervention Styles}
\label{sec: 7.2}

Telling, Selling, and Participating produced comparable objective outcomes (Table \ref{tab:descriptive}), likely because each style carries distinct pros and cons.
Telling (high-task, low-relationship) provide clear and explicit guidance. Yet its authoritative delivery often made participants feel \textit{“managed by a bot”} (P56) or \textit{“interviewed”} (P17), prompting resistance.
Participating (low-task, high-relationship) preserved equality and peer-like rapport, fostering willingness to respond. However, its subtle guidance sometimes failed to provide needed directional input, resulting in stagnation.
This trade-off parallels findings in group leadership research: high-directive styles can compromise perceived autonomy and intrinsic motivation \cite{deci1987support, lewin1939patterns}, while purely relationship-oriented approaches risk under-scaffolding, leading to reduced coordination and goal alignment \cite{house1996path, yukl2012leadership}. 

Selling (high-task, high-relationship) has a great potential to reconcile this tension - preserving accessibility while providing structural guidance. 
Literature suggests that self-explanatory directives can reduce hierarchy perception and strengthen legitimacy \cite{guerin1994analyzing}, and that suggestions, rather than prescriptions, can support autonomy while sustaining progress \cite{soller2001supporting, dillenbourg2002over}.
However, our experiment revealed that Selling did not fully realize this promise. Its \textit{“guiding teacher”} tone was indeed perceived as more personable than Telling’s \textit{“authoritative boss”}, and its persuasive explanations made its instructions more acceptable. But as mentioned in Section \ref{sec:6.2.2}, participants felt an implied obligation, which increased emotional burden and sometimes triggered avoidance. This echoes emotional coercion mechanisms where obligation is used to gain compliance \cite{newman1997emotional}, and when such tactics are perceived, people may exhibit heightened psychological reactance, reduced trust in the source, and engage in withdrawal or resistance to protect autonomy \cite{brehm1966theory}— patterns evident in our observations.
This backfire effect may also stem from participants’ perceptions of AI’s social roles. Even for Selling, the agent implicitly positioned itself as a superior to the human in the interaction, leading some participants (12/60) to ask, \textit {“Why should an AI instruct me?”} This reflects a mismatch between the role AI adopted in the study and human expectations, which appeared to trigger instinctive resistance. As AI’s roles in society continue to evolve, selling-style interventions may gain acceptance in contexts where such authority feels legitimate.

To address this tension, designers may draw on the principle of SLM, which posits that there is no “best” style, and optimal influence is achieved by aligning behaviors with task demands and follower readiness \cite{hersey1979situational}.
In our context, task demands vary across phases of knowledge co-construction, with differing information needs and engagement patterns. 
Therefore, a promising approach is to \textbf{adapt the AI agent’s style dynamically to the evolving phase of knowledge co-construction}, mitigating the limitations of any single style while leveraging its strengths when most advantageous. Our phase-specific insights (see Section \ref{sec:6.2.3}) can serve as an empirical foundation for this strategy.
A potential trade-off in style adaptation is loss of a consistent agent persona. Frequent shifts in communicative style could challenge users’ ability to form stable expectations about the agent, potentially reducing trust or perceived reliability \cite{lee2005can, waytz2014mind}. 
One complementary solution is a \textbf{multi-agent intervention framework}, in which specialized agents maintain distinct, internally coherent styles aligned with their facilitation role and intervene only in phases where their style is most effective. In some cases, this design can enable parallel and complementary interventions. For example, when a discussion needs to transition into the negotiation phase, a selling-style agent can prompt participants to initiate negotiation around those discrepancies, and a participating-style agent can respond by adopting a clear position to reconcile differing viewpoints. Operating in parallel, the agents might generate synergistic effects and better address the heterogeneous needs of users \cite{traum2008multi,bos2002effects}.
Such multi-agent design can reduce users' cognitive load of interpreting fluctuating agent behaviors, allowing each agent to build rapport and credibility within its defined scope of responsibility \cite{bos2002effects, chaves2021should}. 

Regarding the follower readiness dimension of SLM, while participants are typically fluid in asynchronous platforms, it is still applicable in platforms that support private messaging. By drawing on interaction history to estimate a user’s contribution ability, motivation, and receptivity to different styles, the agent could \textbf{deliver personalized interventions through direct messages.}

\subsection{Other Design Considerations}

\subsubsection{Attention is a scarce resource: AI should serve as glue, not gravity}
\label{sec:7.3.1}

Participants reported two contrasting attentional effects of the agent on human–human interaction. Acting as \textit{glue}, AI amplified the visibility and accessibility of overlooked contributions by elaborating on them, bridging knowledge gaps and fostering resonance among participants who might otherwise disengaged. As \textit{gravity}, however, it drew interactions toward itself; participants replied to AI rather than to one another, particularly when it adopted a directive style. Excessively comprehensive responses further reduced opportunities for others to build upon contributions, weakening human-human ties.

To enhance human-human interaction, intervention design should follow a \textit{one-intention-per-turn} heuristic: deliver a single move or intention at a time. If two intentions are necessary (e.g., seeding an idea and explaining its relevance), split them across turns with time in between. This reduces cognitive load and preserves space for peer contributions, aligning with best practices in facilitating distributed collaboration \cite{dillenbourg2002over, suthers2006technology}. Additionally, message frequency and length should be moderated to avoid monopolizing attention.
Moreover, we suggest continuously monitoring the \textit{interaction edge ratio} -- the proportion of human–human edges relative to human–AI edges -- as a live indicator of interaction balance. A rising dominance of human–AI edges signals a drift toward gravity, warranting shorter, less frequent interventions. Under a phase–style adaptive framework (Section \ref{sec: 7.2}), this metric can also inform style selection: when human–AI edges dominate, task-oriented styles should give way to more participatory approaches.

\subsubsection{Recognition matters: even bot replies reduce the “void”}

A salient finding is that receiving any reply -- human or AI -- can legitimize participation. Several participants noted that even a brief acknowledgment from a bot reduced the sense of \textit{“shouting into the void”} and of being ignored, thereby lowering the perceived social cost of contributing again (P3, P24). This effect aligns with the Computers Are Social Actors theory, which posits that humans tend to anthropomorphize responsive systems, applying interpersonal norms to their interactions \cite{nass1994computers}. Therefore, when an agent replies, it is often perceived as a socially meaningful response that affirms the contributor’s presence. Such acknowledgment can activate reciprocity norms common in online communities, which in turn encourage re-engagement \cite{tidwell2002computer, lampe2004slash}.

Such recognition may be particularly valuable for novice contributors, who experience greater uncertainty and social risk when entering a discussion \cite{lampe2004slash}. AI agents can prioritize detecting contributions from newcomers and selectively offering validation or praise, creating low-cost, high-impact interventions to foster continued participation. However, excessive or indiscriminate recognition should be avoided, as it may shift conversational gravity toward the AI (Section \ref{sec:7.3.1}), diminish the perceived value of approval, weaken perceived sincerity, and potentially undermine intrinsic motivation \cite{deci1999meta, henderlong2002effects} . 

\subsubsection{Shifting synthesis: from manual production to assisted refinement}
\mxj{
The result indicates that expecting users to independently synthesize knowledge in Phase 4 is often unrealistic, although workshop participants suggested that active synthesis and contributor agency are critical for solidifying consensus (Section \ref{design-workshop}). We observed that directive interventions (Telling/Selling) largely failed because participants perceived summarization not merely as labor, but as an act of claiming authority over the group's knowledge. Consequently, users are inclined to relinquish agency in this phase when encountering these cognitive and social barriers. 
However, fully automating this phase risks reducing contributors to passive consumers, bypassing the cognitive engagement required to consolidate a shared understanding \cite{fishkin2009people}. We therefore propose a design shift from \textit{human production} to \textit{human refinement and reflection}. Instead of asking users to synthesize from scratch, agents should act as "draftsmen," generating an initial summary. This approach lowers cognitive barriers and mitigates social pressure \cite{heritage2012epistemics, heritage2005terms}: users are no longer "claiming authority" over peers but simply refining and reflecting upon an AI's artifact. Thus, user agency is preserved through the higher-order tasks of validation and reflection—ensuring the consensus remains human-verified while offloading the labor of formulation.
}

\subsubsection{Ethical Concerns}

\

\textbf{Responsible Use of LLMs in Knowledge Communication:}
LLMs are advancing rapidly but remain prone to “hallucinations,” producing plausible yet inaccurate information \cite{ji2023survey}. In one-to-one knowledge exchange, such misinformation can be highly detrimental; however, in knowledge co-construction settings, its impact may be less severe. Online communities often exhibit robust self-regulation mechanism, where participants collectively identify and correct inaccuracies, thus mitigating the spread of misinformation \cite{yuan2025love, zhang2025coknowledge}. In some cases, such collaborative misinformation correction can deepen discourse from exploration toward negotiation \cite{he2021beyond}. Nonetheless, when designing LLM-based agents as active community members, researchers should adopt a cautious and transparent approach, making users aware of the potential for error and the inherent limitations of LLM-generated content.

\textbf{Minimizing Intrusiveness in Personal Probing:}
As discussed in Section~\ref{sec:UX}, when agents ask increasingly specific questions about sensitive, identifiable information—such as participants’ real-life occupations—users may perceive the interaction as overly intrusive, leading to discomfort and anxiety. To safeguard trust and address privacy concerns, researchers should design interaction protocols that limit the depth of personal probing and ensure transparency about why such information is requested and how it contributes to the collective knowledge-building process.


\textbf{Preserving Human Role in AI-Supported Knowledge Co-Construction:}
During the interview, P32 stated that the participating-style agent appeared \textit{“too intelligent,”} expressing uncertainty about whether such agents might eventually replace all human members, leaving AI entities to converse exclusively with one another in the co-construction of knowledge. This highlights an ethical concern: although AI agents can enhance the depth of discussions, they may also lead participants to fear that their own contributions will be marginalized, thereby reducing their motivation to engage and undermining the social and educational value of active human involvement. To sustain human motivation and perceived contribution, designs could incorporate transparent source attribution for agent-generated knowledge, especially when it derives from public, human-authored materials. By explicitly acknowledging these origins, the system can convey that its outputs are grounded in existing human knowledge rather than originating solely from the AI itself. Additional approaches may involve deliberately leaving certain knowledge gaps to prompt human participants to contribute, thereby sustaining their active role in the co-construction process.

\subsection{Limitation and Future Work}


This study has several limitations. \textbf{First}, our within-subject laboratory design—where each participant engaged in one discussion thread per day within a fixed six-person group over five consecutive days—enabled tight experimental control, minimized data contamination, and partially simulated authentic knowledge co-construction processes. 
However, this setting does not fully capture real-world dynamics, where participants of a thread are fluid throughout the day, and individuals often join multiple threads per day. 
Moreover, while our threads resembled small- to medium-scale discussions on real platforms, large-scale threads with thousands of comments may exhibit different interaction patterns and knowledge-building dynamics.
Therefore, while the present findings provide valuable evidence, future research should conduct larger-scale, longer-term field studies in fully naturalistic contexts.
\textbf{Second}, the participant sample consisted predominantly of young adults (ages 18–31) from East Asian cultural backgrounds. Further investigations should extend to participants of more diverse ages and cultural backgrounds.
\textbf{Third}, to ensure that all participants could meaningfully contribute, we selected accessible, everyday discussion topics. While appropriate for our recruitment scope, such topics may elicit less “hardcore” or technically specialized knowledge exchange compared to scientific topics. Future research should therefore investigate the effectiveness of agent interventions in more knowledge-intensive subject contexts. 
\textbf{Fourth}, our interface adopted a chronological layout; alternative designs, such as tree-structured layouts, may shape discussion dynamics differently. Future research should compare agent effects across platforms with varied front-end interfaces.
\mxj{
\textbf{Fifth}, we acknowledge the potential for a demand effect \cite{orne2017social, nass2000machines}, where participants feel compelled to “play along” with the agent. Our study did not measure this effect directly; future work should evaluate it in more naturalistic contexts where users have full autonomy to ignore interventions, thereby distinguishing such compliance from users’ genuine preferences.
}


\section{Conclusion}

In this paper, we propose a process-orchestrated intervention paradigm for AI agents that progressively advance knowledge co-construction in asynchronous online discussions. The specific intervention strategies for task-oriented and relationship-oriented styles within this paradigm were derived from a design workshop. Implementing these strategies in an LLM-based agent, we conducted a within-subject study to examine their potential impact on knowledge co-construction dynamics, user perceptions and experiences, and human–human interactions. 
Our results demonstrate the promise of such intervention logic in fostering deeper and more robust collective knowledge. They also reveal the differentiated effects of task- and relationship-oriented styles on the co-construction process and participant behaviors, leading to several design considerations and research directions. 
With these empirical insights, we position this work as an exploratory step toward developing contextual AI agents that more effectively enhance online discourse as an active member.


\begin{acks} 

This project is supported by the Hong Kong SAR Research Grants Council's Theme-based Research Grant Scheme (Project No. T43-518/24-N). We also extend our gratitude to the reviewers for their constructive feedback, which had significantly improved the quality of our paper.

\end{acks}

\bibliographystyle{ACM-Reference-Format}
\bibliography{CHI26}

\end{document}